\documentclass[prd,aps,twocolumn,a4paper,floatfix,showpacs,nofootinbib]{revtex4-1}

\usepackage{graphicx,psfrag}
\usepackage{mathrsfs}
\usepackage{amsmath,amsfonts,amssymb}
\usepackage{multirow}
\usepackage{comment,hyperref}

\newcommand{\be}{\begin{equation}}
\newcommand{\ee}{\end{equation}}
\newcommand{\bea}{\begin{eqnarray}}
\newcommand{\eea}{\end{eqnarray}}
\newcommand{\bel}{\begin{align}}
\newcommand{\eel}{\end{align}}
\newcommand{\vecb}[1]{\mathbf{#1}}

\def\p{\partial}

\def\half{\frac{1}{2}}

\def\e{{\rm e}}
\def\i{{\rm i}}

\def\GMc2{{\rm G M_{\odot} c^{-2}}}

\def\O{\mathcal{O}}

\def\ipoh{{i+1/2}}
\def\imoh{{i-1/2}}
\def\check{$\checkmark$}
\def\cross{$\times$}
\def\Rich{\mathcal{R}}

\usepackage{color}

\definecolor{cyan}{rgb}{0,0.9,0.9}
\definecolor{orange}{rgb}{0.9,0.5,0}
\definecolor{magenta}{rgb}{1,0,1}
\definecolor{purple}{rgb}{0.8,0.4,0.8}
\definecolor{gray}{rgb}{0.5,0.5,0.5}

\begin{document}

\title{Gravitational waveforms from binary neutron star mergers \\
with high-order WENO schemes in numerical relativity} 

\author{Sebastiano \surname{Bernuzzi}${}^{1}$}
\author{Tim \surname{Dietrich}${}^{2}$}
\affiliation{${}^1$DiFeST, University of Parma and INFN Parma, I-43124 Parma, Italy}
\affiliation{${}^2$Max-Planck-Institut for Gravitational Physics, Albert-Einstein-Institut, D-14476 Golm, Germany}

\date{\today}

\begin{abstract}
The theoretical modeling of gravitational waveforms from binary
neutron star mergers requires precise numerical relativity simulations. 
Assessing convergence of the numerical data and building the error budget
is currently challenging due to the low accuracy of
general-relativistic hydrodynamics schemes and to the grid resolutions
that can be employed in $(3+1)$-dimensional simulations.    
In this work, we explore the use of high-order
weighted-essentially-nonoscillatory (WENO) schemes in neutron star
merger simulations and investigate the
accuracy of the waveforms obtained with such methods.
We find that high-order WENO schemes can be robustly employed for
simulating the inspiral-merger phase and they significantly improve
the assessment of the waveform's error budget with respect to
finite-volume methods.  
High-order WENO schemes can be thus efficiently used for high-quality
waveforms production, also in future large-scale  
investigations of the binary parameter space. 
\end{abstract}


\maketitle

\section{Introduction}
\label{sec:intro}

Gravitational-wave (GW) astronomy has started with the
first direct observation of a merging binary black hole system 
\cite{Abbott:2016blz}. Beside black hole binaries, binary neutron
stars (BNS) are one of the main expected sources for the advanced GW
detectors \cite{Aasi:2013wya}. The BNS parameters estimation, 
including the NS's equation of state, in near-future detection
requires models of {\it merger} waveforms. An example of such a model is
the tidal effective-one-body model we have recently developed in
\cite{Bernuzzi:2014owa}, which is compatible with numerical relativity
(NR) data up to merger (peak of the waveform's amplitude)
\cite{Bernuzzi:2014owa}. Developing and improving the accuracy of
these analytical models crucially relies on the availability of precise NR
waveforms. This paper reports on our latest 
effort towards the production of NR waveforms from BNSs.  

Key issues for producing high-quality NR waveforms are (1) the assessment of
the convergence properties of the numerical data at increasing grid
resolutions and (2) the production of a trustworthy error budget. 
Different NR groups have considered this
problem~\cite{Baiotti:2009gk,Bernuzzi:2011aq,Bernuzzi:2012ci,Reisswig:2012nc,Radice:2013hxh,Hotokezaka:2013mm,Bernuzzi:2014owa,Hotokezaka:2015xka,Haas:2016cop} 
and showed that 3+1 NR simulations including
general-relativistic hydrodynamics (GRHD) can deliver waveforms with
phase uncertainties of
$\delta\phi\lesssim3$~rad \cite{Radice:2013hxh,Bernuzzi:2014owa,Hotokezaka:2015xka}  
accumulated over the last $\sim10-20$ orbits to merger~\footnote{The 
  phase $\phi$ of the complex GW $h=A\,\e^{ -\i\phi}$ is the most
  relevant quantity computed in the simulations and the most sensitive
  to numerical errors.}.  
These are rather accurate results but, on the one hand, convergence
and error bars are typically difficult to estimate (see below), and, on
the other hand, the improvements of the current analytical models 
require computations at higher precision \cite{Bernuzzi:2014owa}.

All previous works agree on that the truncation error of the
GRHD solver is the main source of uncertainty in BNS simulations,
e.g.~\cite{Bernuzzi:2011aq,Hotokezaka:2013mm,Radice:2013hxh}. 
The numerical dissipation of the second-order-accurate 
shock-capturing algorithms employed for the
solution of GRHD, combined with the typical 3D grid resolutions, can 
dramatically affect the accuracy of the phase evolution of the
binary~\cite{Bernuzzi:2012ci}. 
The use of nonlinear limiters in shock-capturing schemes
introduces systematic errors that are difficult to
quantify~\cite{Bernuzzi:2012ci}. 
Furthermore, measuring the convergence
rate of the numerical solution is challenging because of the limited
span in grid resolutions that can be achieved in 3D
simulations~\cite{Bernuzzi:2011aq}. And in absence of a clear
convergence 
measure, the evaluation of truncation errors is impossible. 
Improving the numerical schemes for GRHD is thus necessary for future
development of the field.

In a series of seminal work in the 1990s
\cite{Harten:1987,Harten:1989,Shu:1988,Shu:1989,Liu:1994,Jiang:1996}, 
Harten, Shu and others developed high-order finite-differencing
flux-conservative schemes for hydrodynamics. These schemes rely on
three elements: projection of the fluxes on characteristic fields, 
the flux-splitting approach, and high-order--essentially-nonoscillatory
interpolation procedures (reconstruction). A particularly efficient
reconstruction  is the fifth-order
weighted-essentially-nonoscillatory (WENO) interpolation scheme
of~\cite{Jiang:1996}, which is very well known in the computational physics literature.
Such WENO high-order algorithms are natural candidates for NR codes
and BNS evolutions for several reasons:  
(i)~they can achieve high-order accuracy for sufficiently smooth solution; 
(ii)~they are developed for uniform grids and Runge-Kutta (RK) time
stepping, which are both standard techniques in NR;  
(iii)~they are consistent with the finite-differencing treatment used
for the metric fields;  
(iv)~and they can be easily implemented in multidimensions. 

Radice et al. presented, for the first time in the context of NR, BNS merger
simulations employing a high-order finite-differencing algorithm \cite{Radice:2013hxh,Radice:2013xpa,Radice:2015nva,Radice:2016gym}. 
The latter is very similar to the WENO method mentioned above, the
only difference being the use of the fifth-order monotonicity preserving (MP5)
scheme of Suresh et al.~\cite{Suresh:1997}. 
BNS simulations benefit from the use of the high-order scheme. In
particular, third-order convergence was observed for the waveform's
phase using four simulations at different resolutions
\cite{Radice:2015nva,Radice:2016gym}. The MP5-based high-order 
scheme has been also recently used for the simulations
presented in \cite{Hinderer:2016eia}, although no detailed 
convergence analysis has been presented there. 

In this work we assess, for the first time, the use of
high-order WENO schemes for the computation of GWs from BNS
inspiral-merger simulations. We focus on the accuracy of the GW phase
error and show that, although the scheme does not achieve the formal
high-order accuracy, convergence can be clearly
monitored at typical 3D grid resolutions. 
The error budget is then robustly determined by combining numerical
data from different grid resolutions and using Richardson extrapolation.
We also consider the MP5 scheme~\cite{Suresh:1997} for comparison with
previous work, but we do not find this method as robust as the WENO
one in our simulations.

The paper is organized as follows.
Sec.~\ref{sec:eqs} summarizes the relevant equations.
Sec.~\ref{sec:num} describes the numerical methods; and it is
complemented by Appendix~\ref{app:flat_test} in which our implementation is
validated against standard test problems.
Sec.~\ref{sec:ss} describes numerical evolutions of single star
spacetimes with the high-order scheme. 
Sec.~\ref{sec:bns} describes numerical evolutions of inspiral-merger
BNS with the high-order WENO scheme. Focus is on convergence of the
data for increasing grid resolutions.
Sec.~\ref{sec:err} discusses error estimates for our BNS waveforms. 
Throughout this work we use geometric units setting $c=G=1$ and masses
are expressed in $M_\odot$.

\section{GRHD and Dynamical Spacetimes}
 \label{sec:eqs}

In this section we briefly review the equations governing relativistic  
flows on dynamical spacetimes in 3+1 NR, as used in this work. 
The problem is defined by a PDE system composed of GRHD in
conservation form~\cite{Font:2007zz},    
\be
\label{hydro}
\partial_t \vecb{q} = - \partial_i \vecb{f}^{(i)} + \vecb{s} \ ,
\ee
coupled to a 3+1 hyperbolic formulation of the metric equations, schematically
\be
\label{metric}
\partial_t \vecb{u} = {\cal{N}} 
\left(\partial^2_i \vecb{u},\partial_i\vecb{u}, \vecb{u};\vecb{w}\right) \ . 
\ee
In Eq.~\eqref{hydro}, $\vecb{q}$ is a state vector collecting the
conserved variables, $\vecb{f}$ are the hydrodynamical fluxes, and
$\vecb{s}$ is a source term depending on metric fields and primitive
variables.
The conserved variables are $\vecb{q} = \sqrt{\gamma}(D,\,
S_k,\,\tau)$, and represent the rest mass density ($D$), the momentum
density ($S_k$), and an internal energy ($\tau$) of the Eulerian
observers that define the spacetime foliation. The quantity
$\gamma=\det\gamma_{ij}$ is the determinant of the spatial metric in
the 3+1 decomposition of Einstein equations.
Conserved variables can be written in terms of the primitive
variables $\vecb{w}=(\rho,v^i,\epsilon, p)$, i.e.~rest mass density,
3-velocity, internal energy, and pressure of the fluid 
(comoving frame). An equation of state (EOS) closes 
the system by specifying the pressure in terms of the density $\rho$ and
internal energy $\epsilon$. In this work, the neutron star matter is described by
either a $\Gamma$-law EOS, $p = (\Gamma-1)\rho\epsilon$, 
or a more realistic, zero temperature, 
SLy and MS1b EOS~\cite{Douchin:2001sv}. Isentropic
evolutions of $\Gamma$-law EOS models are sometimes imposed by
specifying a polytropic EOS, $p = K\rho^\Gamma$, $K=const$ and without
evolving the GRHD equation for $\tau$, which is then redundant. 
The zero temperature EOS are 
implemented by a piecewise polytrope fit~\cite{Read:2008iy}, and
thermal effects are modeled by an additive pressure
contribution given by the $\Gamma$-law EOS with
$\Gamma=1.75$~\cite{Shibata:2005ss,Bauswein:2010dn,Thierfelder:2011yi}.  

Let us briefly comment on the metric equations. In Eq.~\eqref{metric},
$\vecb{u}$ is a state vector collecting the 
components of the tensorial metric variables, and the right-hand-side
(r.h.s)  
operator ${\cal N}$ is a nonlinear function quadratic in the field variables and first
derivatives. As common in 3+1 numerical relativity, we employ a free
evolution approach to Einstein equations. During the evolution of
constraint satisfying (consistent) initial data, Einstein equations
can be violated at the numerical error level.
The choice of an appropriate formulation of general relativity is a
key point for controlling and reducing these violations.
In this work, we use the Z4c formulation proposed in~\cite{Bernuzzi:2009ex} (see
also~\cite{Hilditch:2012fp}) and the BSSNOK
system~\cite{Nakamura:1987zz,Shibata:1995we,Baumgarte:1998te}.
The system in Eq.~\eqref{metric} contains evolution equations for
the gauge, i.e.~the lapse and shift evolution equations. We use the moving
puncture gauge implemented as described
in~\cite{Brugmann:2008zz,Thierfelder:2010dv,Hilditch:2012fp}.

\section{Numerical Framework}
\label{sec:num}

\subsection{Grid tructure and evolution algorithm}
\label{sec:gen}

Our simulations are performed with the BAM
code~\cite{Dietrich:2015iva,Thierfelder:2011yi,Brugmann:2008zz}.  
The equations described above are numerically solved by explicit time
evolution of the initial data.
The evolution algorithm is based on the method of lines with explicit
RK time integrators. A portion of
the spacelike hypersurfaces $\Sigma_t$ is covered by logically Cartesian overlapping
grids. 
The strong-field region is covered by a hierarchy of
cell-centered and nested Cartesian grids.
The hierarchy consists of $L$ levels of refinement labeled
by $l = 0,...,L-1$. A refinement level consists of one or more
Cartesian boxes with constant grid spacing~$h_l$ on level~$l$. A
refinement factor of 2 is used such that~$h_l = h_0/2^l$. The grids
are properly nested in that the coordinate extent of any grid at
level~$l$, $l > 0$, is completely covered by the grids at level~$l-1$.
Some of the mesh refinement levels $l>l^{mv}$ can be dynamically moved
and adapted during the time evolution.
We use $n$ points per direction per fixed level, and $n^{mv}$ points 
per direction per moving level. 
For some cases, we cover the wave zone $l=L-1$ by a cube-sphere
multipatch grid with $n_r$ point in the radial direction and
$n_{\theta,\phi}\sim n/2$ in the angular ones. 

The grid is evolved in time using the Berger-Oliger
algorithm~\cite{Berger:1984zza}. We employ fourth- and third-order RK with a
Courant-Friedrich-Lewy (CFL) factor of~$0.25$ in all the neutron star
tests. (For some simulations we tested a lower CFL condition, but see
no significant change in the convergence behavior.)  
Six buffer zones per side per direction are employed for the Berger-Oliger algorithm
interpolations. The restriction and prolongation is performed with a 
fourth-order WENO scheme for the matter fields, and a sixth-order
Lagrangian scheme for the metric fields.
Contrary to our latest simulations, no additional correction step is
employed here for the conservative variables~\cite{Dietrich:2015iva}. The
chosen grid setup already guarantees excellent mass conservation during the
inspiral-merger. 

The metric field derivatives in~\eqref{metric} are approximated by
fourth-order accurate finite-differencing stencils. Sixth-order artificial
dissipation operators are employed to stabilize noise from mesh
refinement boundaries. The high-order scheme implemented for GRHD is
described below. In our previous work, GRHD equations were
solved with a second-order scheme composed of the local Lax-Friedrich
(LLF) scheme for the fluxes, and primitive 
reconstruction. We found that the WENOZ reconstruction, among several
tested, is particularly accurate for neutron star
evolutions~\cite{Bernuzzi:2012ci}. Hence, it is chosen here for
comparison with the high-order scheme.

In order to simulate vacuum regions, a static, low-density, and cold
atmosphere is added in the vacuum region outside the
star~\cite{Thierfelder:2011yi}. The atmosphere density is chosen as
\be
\rho_{atm} = f_{atm}\max\rho(t=0)\ ,
\ee
and grid points below a threshold $\rho_{thr}= f_{thr}\rho_{atm}$
are set to $\rho_{atm}$. 
As we see, the low-density flow is one of the main sources of error in the neutron
star simulations (see also e.g.~\cite{Radice:2013xpa} for discussions). In
this work, we did not attempt to modify our standard atmosphere
prescription because we aimed at comparing the higher-order flux scheme with our
current ``best'' second-order scheme. Note that {\it all} the numerical relativity
implementations make use of similar assumptions and algorithms at low
densities as those employed here. In general, it is challenging to
deal with matter/vacuum interfaces in the presence of gravitational
fields, even without the complication of dynamical spacetimes. See, 
however, \cite{Xing:2012} for an attempt in the case of a static and
Newtonian gravitational field.

\subsection{High-order finite-differencing schemes}
\label{sec:howeno}

In this section we briefly review the high-order (HO) finite-differencing
scheme~\cite{Shu:1989,Liu:1994,Jiang:1996,Mignone:2010br}.
The presentation is restricted to 1D without loss of
generality: a multidimensional scheme is simply obtained by
considering fluxes in each direction separately and adding them in 
the r.h.s.. We assume a 1D {\it uniform} mesh $x_i$ with $i=1,...,n$
and spacing $h$. Cell interfaces are indicated with $x_\ipoh=x_i + h/2$
and the pointwise values of a function $f_i=f(x_i)$.
Following~\cite{Shu:1989} the divergence term in \eqref{hydro} is
calculated using the conservative finite-differencing formula 
\be
\frac{\partial f}{\partial x}(x_i) = h^{-1}
\left(\hat{F}_\ipoh-\hat{F}_\imoh\right) \ , 
\ee
where the numerical fluxes at the interfaces, $\hat{F}_\ipoh$, are a
high-order nonoscillatory approximation of the so-called \textit{numerical
flux function}. The approximation is computed using pointwise values
$f_i$, identifying them as the cell-averages of another function, and
applying reconstruction via primitive function~\cite{Shu:1989}.

The numerical fluxes for the GRHD system are built using a
flux-splitting approach based on the LLF 
and performing the reconstruction on the characteristic
fields~\cite{Jiang:1996,Suresh:1997,Mignone:2010br}. Specifically, the
characteristic 
fields are given by projections of the positive and negative part of
the flux onto the left eigenvectors matrix of the Jacobian
$\p\vecb{f}/\p\vecb{q}$ computed at $\ipoh$. The $k$th characteristic
fields read (cf. Eq.~(17) of~\cite{Mignone:2010br}) 
\be
\label{charfields}
\hat{F}^{(k)\pm}_{(\ipoh), S} = \half \vecb{L}^{(k)}_\ipoh
\cdot \left(\vecb{f}_{S} \pm a^{(k)} \vecb{q}_{S}\right) \ ,
\ee
where $\vecb{L}^{(k)}_\ipoh$ is the $k$th left eigenvector matrix
computed at $\ipoh$, ${}^\pm$ indicate the positive and negative flux, $S$ is an
appropriate stencil to be used in the reconstruction (see below), and
\be
a^{(k)}=\max_{S}(|\lambda^{(k)}|)
\ee
are the maximum of the absolute values $k$th characteristic speeds on
$S$.   
The intercell characteristic fields are obtained applying a 
reconstruction algorithm, 
\be
\hat{F}^{(k)\pm}_\ipoh = \mbox{Rec}\left[\hat{F}^{(k)\pm}_{(i+1/2),S}\right] \ ,
\ee
here indicated by the operator $\mbox{Rec}[.]$. The latter uses the
characteristic fields on the stencil $S$ [e.g.~Eq.(4.2) of~\cite{Jiang:1996}].
The numerical fluxes are finally obtained projecting back the
characteristic fields
\be
\hat{\vecb{F}}_\ipoh = \sum_{k} \left(
  \hat{F}^{(k)+}_\ipoh+\hat{F}^{(k)-}_\ipoh\right)
\vecb{R}^{(k)}_\ipoh \ .
\ee
The left (right) $\vecb{L}^{(k)}_\ipoh$ ($\vecb{R}^{(k)}_\ipoh$)
eigenvectors at $\ipoh$ are computed from arithmetic averages of $i$
and $i+1$ values. 
Their expression for GRHD can be found
in~\cite{Font:2007zz}. We compute averages of primitives (except pressure)
and metric functions only, and compute the other quantities from these
averages. Note that other, nonequivalent procedures for computing the
averaged eigenvectors are possible. We have checked some of them and
found no differences in the tests of Appendix~\ref{app:flat_test}.     

The HO scheme above is completed by specifying a reconstruction
procedure. Here, we consider the fifth-order
WENOZ~\cite{Borges20083191} and MP5~\cite{Suresh:1997} algorithms
following the implementation of~\cite{Mignone:2010br}. 
A fifth-order WENO uses five points stencil $S^+= (i-2,...,i+2)$ to
reconstruct the grid function $f_i$ at the
interface~\footnote{$f_{\imoh}$ is simply given by shifting each index
  by $-1$ in the $f_{\ipoh}$ formula. In the code we use a single
  routine applied either to the stencil $S^+$ or $S^-= 2 i - S^+ + 1$ \cite{Mignone:2010br}. }, 
\begin{align}
&f_{\ipoh} = \mbox{Rec}\left[f_{S^+}\right] = 
\frac{\omega_0}{6}\left(2f_{i-2}-7f_{i-1}+11f_i\right) \\
&+ \frac{\omega_1}{6}\left(-f_{i-1}+5f_{i}+2f_{i+1}\right)
+ \frac{\omega_2}{6}\left(2f_{i}+5f_{i+1}-f_{i+2}\right) \nonumber \ .
\end{align}
The weights are nonlinear functions
$\omega_j=\alpha_j/\sum_{j=0}^2\alpha_j$ of smoothness indicators
$\beta_j$, and they sum up to unity $\sum_j\omega_j=1$. For WENOZ one
has~\cite{Borges20083191}  
\be
\alpha_j =
o_j\left(1+\frac{|\beta_0-\beta_2|}{\beta_j+\varepsilon}\right) \ ,
\ee
with $\varepsilon=10^{-42}$, optimal weights $o_j = (1/10,6/10,3/10)$
corresponding to the Lagrangian five points interpolation, and
\begin{subequations}  
\begin{align}
 \beta_0 &= \frac{13}{12}  ( f_{i-2} - 2 f_{i-1} + f_i)^2 
 + \frac{1}{4} (f_{i-2}-4 f_{i-1} + 3 f_i)^2\\
 \beta_1 &= \frac{13}{12} ( f_{i-1} -2 f_i + f_{i+1})^2 
+ \frac{1}{4} (f_{i-1}-f_{i+1})^2\\
 \beta_2 &= \frac{13}{12} ( f_i - 2 f_{i+1} + f_{i+2})^2 
+ \frac{1}{4} (3 f_i - 4 f_{i+1} + f_{i+2})^2 
\end{align}
\end{subequations}  
Smoothness indicators are derived from cell-averages of the
derivatives of the interpolating polynomials
\cite{Jiang:1996}. Flatter reconstructed profiles inside the cell give
smaller $\beta_j$ values.

The convergence properties of WENO schemes have been studied in
detail~\cite{Henrick2005542,Tchekhovskoy:2007zn,Borges20083191}. 
WENO schemes combine reconstructions obtained
by different substencils of $S^+$ using the weights
$\omega_k$. Locally, a sufficiently smooth 
solution is reconstructed using the full stencil $S^+$ with the
$\omega_j\sim o_j$ ($\beta_j\to0$). The necessary and sufficient
conditions that the weights have to fullfil in order to achieve the
nominal fifth-order convergence are given in~\cite{Henrick2005542}.
In case of shocks, the substencil containing local
discontinuities or large 
grandients has $\omega_j\to0$ (large $\beta_j$), and the
reconstruction is then performed with the combination of the other
smoother substencils. 

Similarly to WENO, the MP5 reconstruction employs the same stencil and 
the fifth-order Lagrangian interpolation based on that. 
The latter is limited a posteriori according to a
sophisticated monotonicity preserving limiter described
in~\cite{Suresh:1997}.

The BAM implementation of the HO schemes has been validated
using several 1D tests. As summarised in Appendix~\ref{app:flat_test}, the
HO schemes achieve the nominal fifth-order convergence for sufficiently
smooth solutions albeit resolutions of $n\sim1000$ are typically
necessary to clearly measure such convergence properties. 
This suggests that in 3D
applications, as neutron star evolutions, one should 
\textit{not} expect to measure the formal convergence rate at typical
resolutions of $n^3\sim100^3$. The HO schemes also pass standard
relativistic blast wave test, showing shock-capturing properties (see
Appendix~\ref{app:flat_test} for details.)

For the evolution of neutron stars we further implement a ``hybrid''
algorithm that employs the HO 
scheme above a certain density threshold $\rho_{hyb}$ and switch to
our standard 2nd-order LLF method below $\rho_{hyb}$. 
We refer to this algorithm as HO-Hyb.
One of the reasons to perform the switch is that the eigenvector 
matrices of the HO-algorithm contain singular terms for $\rho\to0$.
The density threshold is calculated as 
\be  
\rho_{hyb} = f_{hyb} \ f_{thr} \ \rho_{atm} \label{eq:rho_thr} \ ,
\ee
by introducing the free parameter $f_{hyb}$. As we see in the
next section, this algorithm is more robust than simple HO at low
densities and helps handling the artificial atmosphere.  

Finally, note that the HO algorithm implemented in BAM has already been employed
in~\cite{Bugner:2015gqa} as a subcell algorithm for a discontinuous
Galerkin implementation of GRHD. In that context, it was used for the
evolution of a compact star.

\section{Single Star Evolutions}
\label{sec:ss}

\begin{table}[t]
  \centering    
  \caption{Single star runs. Columns: EOS, numerical flux
  scheme, reconstruction, the atmosphere parameters $f_{atm},f_{thr}$, and the
  parameter $f_{hyb}$ employed in HO-Hyb. All the runs are performed
  at resolutions $n=(48,64,96,128)$.}
  \begin{tabular}{ccccccc}        
    \hline
    \hline
    EOS & Flux & Rec & $f_{atm}$ & $f_{thr}$ & $f_{hyb}$ \\
    \hline
    polytrope & LLF & MP5 & $10^{-11}$ & $10^{2}$ & - \\
    polytrope & LLF & WENOZ & $10^{-11}$ & $10^{2}$ & - \\
    polytrope & HO & MP5 & $10^{-11}$ & $10^{2}$ & - \\
    polytrope & HO & WENOZ & $10^{-11}$ & $10^{2}$ & - \\
    polytrope & HO-Hyb & MP5 & $10^{-11}$ & $10^{2}$ & 5 \\
    polytrope & HO-Hyb & WENOZ  & $10^{-11}$ & $10^{2}$ & 5 \\
    \hline
    $\Gamma=2$ & LLF & MP5 & $10^{-11}$         & $10^{2}$ & - \\
    $\Gamma=2$ & LLF & WENOZ & $10^{-11}$         & $10^{2}$ & - \\
    $\Gamma=2$ & HO & MP5 & $10^{-7}$          & $10^{2}$ & - \\
    $\Gamma=2$ & HO & WENOZ & $5 \times 10^{-8}$ & $10^{2}$ & - \\
    $\Gamma=2$ & HO-Hyb & MP5 & $10^{-11}$         & $10^{2}$ & $2.5 \times 10^4$ \\
    $\Gamma=2$ & HO-Hyb & WENOZ & $10^{-11}$         & $10^{2}$ & $10^4$\\
    \hline
    \hline
     \end{tabular}
  \label{tab:single_star}  
\end{table}

\begin{figure}[t]
  \centering
  \includegraphics[width=0.49\textwidth]{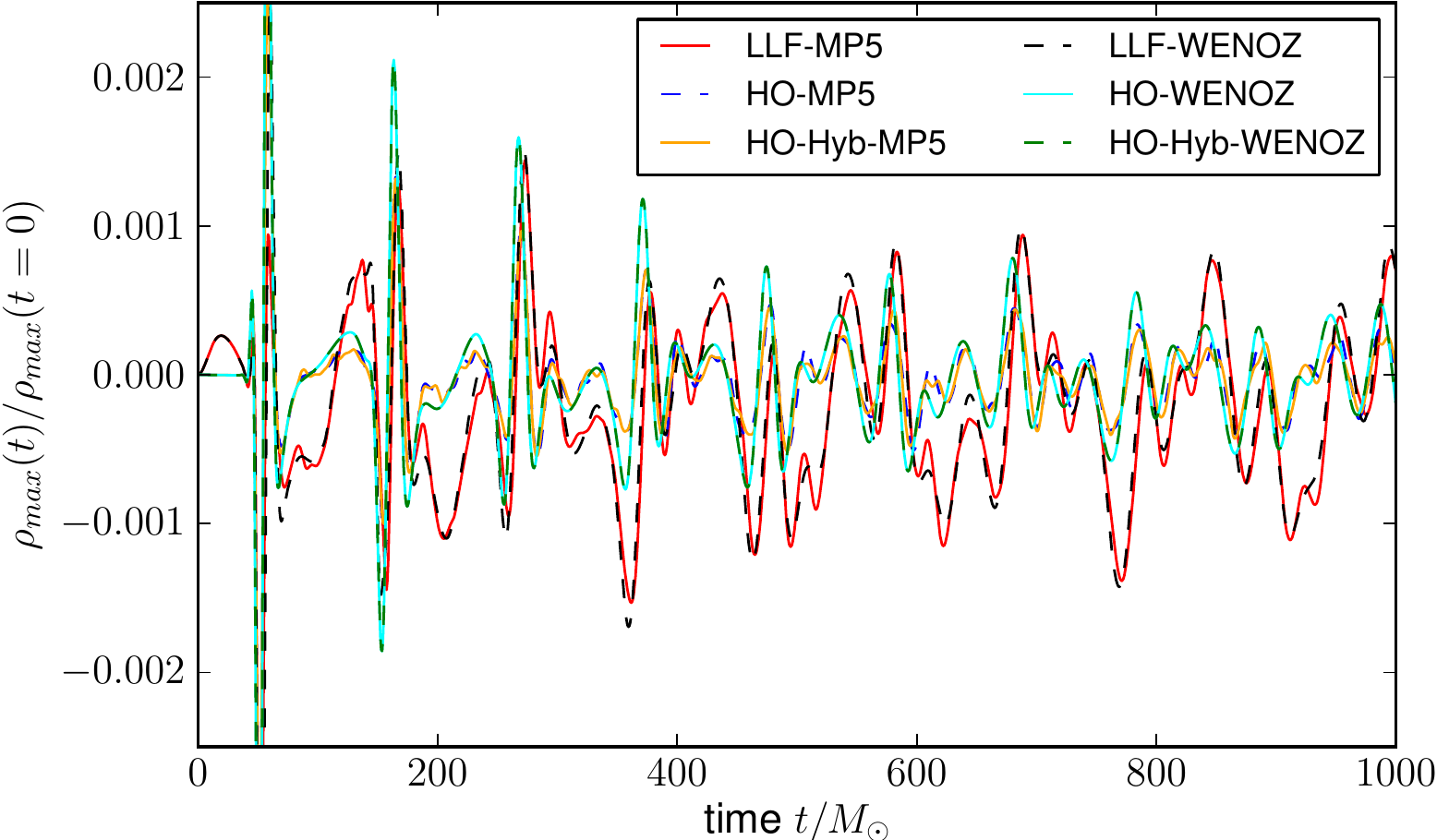}
  \includegraphics[width=0.49\textwidth]{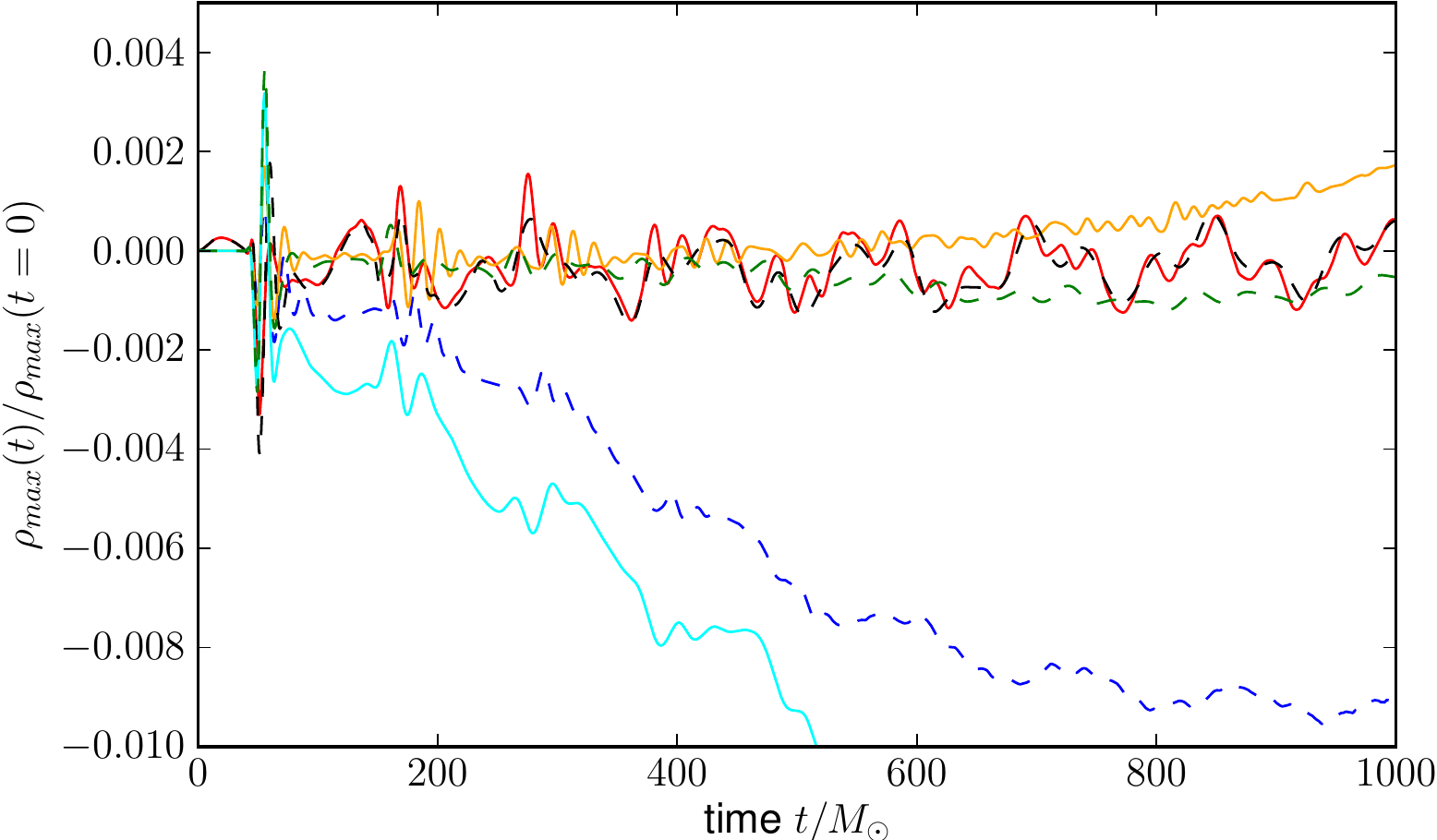}
  \caption{Central rest mass density evolution for simulations of a
  single spherical star with
  $n=96$ and $h=0.1875$. Top: polytropic EOS; Bottom: $\Gamma$-law EOS. }  
    \label{fig:star_rhoc} 
\end{figure}

Evolutions of static star spacetimes are standard benchmarks
for general-relativistic implementations, see
e.g.~\cite{Thierfelder:2011yi,Radice:2013xpa}. 
Here, before presenting BNS evolutions, we assess the convergence of
the HO and HO-Hyb schemes for the star solution, compare different 
reconstruction procedures, and contrast the results with those
obtained with the LLF second-order scheme. 
This benchmark allows one to study many features of generic neutron
star evolution in a simple and controlled setup.

Our tests show that stable long-term evolutions of star
spacetimes can be achieved with HO methods, but convergence is limited
to second-order. For the WENOZ 
reconstruction, we shall show that the origin of the inaccuracy is related
to the specific values of the WENO weights that are nonoptimal
both in the bulk of the matter distribution and at the star
surface. Thus, at typical resolutions for 3D runs, one does not expect
an optimal convergence rate.   

The star initial model is a $\Gamma=2$ polytrope with gravitational
mass $M=1.400$, baryonic mass $M_b=1.506$ and central rest mass
density $\rho_c=1.28\times 10^{-3}$. The BSSNOK evolution scheme is
used for the metric evolution. We employ octant symmetry to reduce the
computational costs. 

The grid is composed of three fixed refinement levels. 
Simulations are performed with $n=(48,64,96,128)$ points covering the neutron star
leading to a grid spacing of $h=(0.375,0.28125,0.1875,0.140625)$. 
It is ensured that the finest box covers the star entirely.
Table~\ref{tab:single_star} summarizes the simulations and settings.
Note that for both the HO and HO-Hyb schemes, the MP5 reconstruction
requires larger values of $f_{atm}$ and $f_{hyb}$ than the WENOZ one
in order to achieve stable long-term evolutions. 

At the continuum level the evolution of a static solution is
trivial. Numerically, if a scheme allows long-term stable evolution, 
discretization errors trigger radial oscillations, which are observable, for
example, in the central rest mass density. The oscillations are
triggered by atmosphere effects and converge to 0 
increasing the resolution. Figure~\ref{fig:star_rhoc} (top panel)
shows the oscillations of the 
central rest mass ($\rho_{\max}$) density over a simulation time of about several
radial periods for the polytropic EOS.  
We compare the LLF, HO, and HO-Hyb scheme 
for both WENOZ and MP5 reconstruction. The figure suggests that the HO
scheme is slightly more accurate than the LLF for all the
reconstructions considered, i.e.~the amplitude of the oscillations is  
smaller. 

The role of the star's surface is illustrated by considering
evolutions with the $\Gamma$-law EOS, bottom panel of Fig.~\ref{fig:star_rhoc}.
In our experience, truncation errors with such EOS are larger than 
the polytropic case with any numerical scheme, because significant unphysical
shock-heating is observed at low-densities.
Indeed, the $\rho_{\max}$ oscillations are larger than in
the polytropic case. The best results for this test are obtained with
the LLF-schemes and the HO-Hyb-WENOZ scheme.  
Moreover, the HO scheme gives by far the
largest discrepancy to the initial configuration. This is due to the
larger atmosphere threshold that have to be employed 
for stable simulations with the HO scheme, see Tab.~\ref{tab:single_star}.

\begin{figure}[t]
  \centering
  \includegraphics[width=0.5\textwidth]{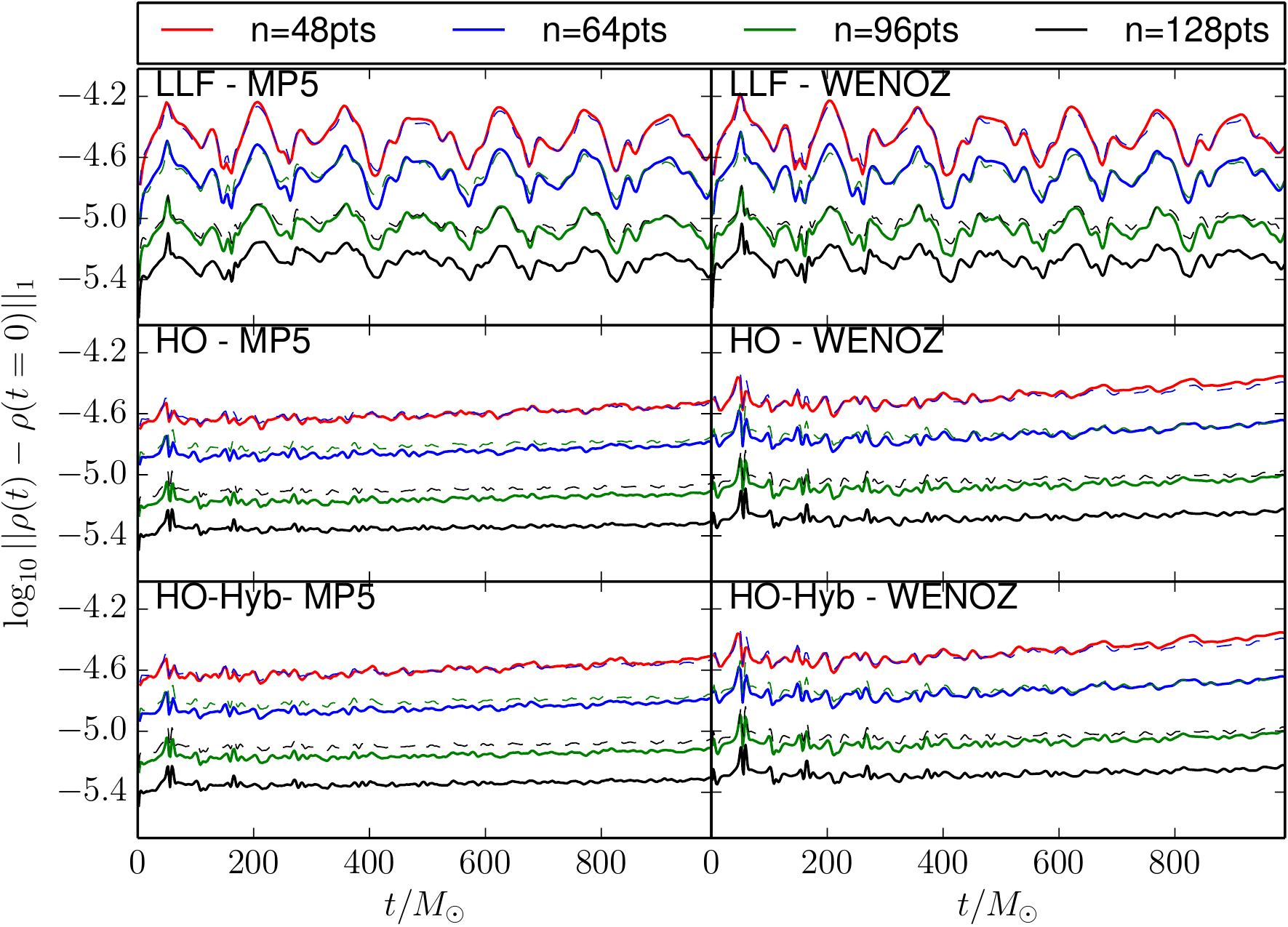}
    \includegraphics[width=0.5\textwidth]{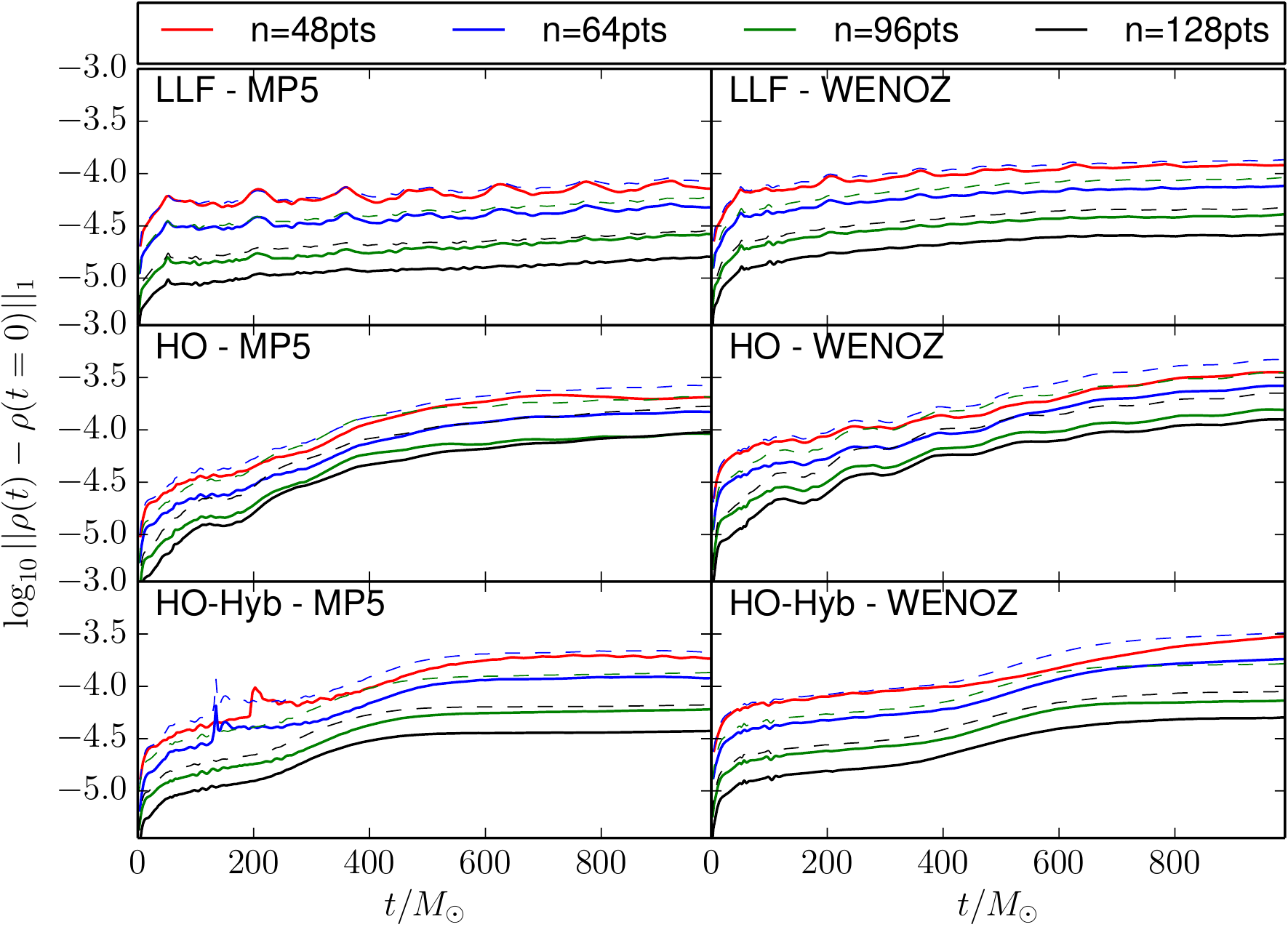}
  \caption{Evolution of the $L_1$-distance $||\rho(t)-\rho(0)||_1$ for
  different numerical methods and resolution. Different resolutions
  are shown with different colors. The dashed colored lines are the same for
  the corresponding solid but 
  scaled according to second-order convergence. 
  Top: polytropic EOS. Bottom: $\Gamma$-law EOS.} 
  \label{fig:star_l1_fgr}
\end{figure}

In order to verify the convergence rate of the scheme we consider the
$L_1$ norm of the difference between 
the evolution profile and the initial data (exact
solution). Figure~\ref{fig:star_l1_fgr} plots the $L_1$ distance
from the exact solution for all resolutions and schemes for the polytropic (top) and $\Gamma$-law
(bottom) EOS. 
Focusing on the top panel, we observe that the absolute error of the
HO and HO-Hyb schemes is smaller and less oscillatory than that of the LLF.
Also, the oscillation amplitude in the norm is smaller. The
convergence rate of \textit{all} the schemes is approximately second
order (cf. dashed lines).
The bottom panel shows that, at the same resolutions, the absolute
errors in the $\Gamma$-law evolutions are approximately ten 
times larger. In this case, it is more difficult to maintain 
the initial equilibrium configuration. The larger
deviations in the norms are \emph{not} due to mass losses,
but mainly to the
deformation of the initial (exact) star profile due to truncation errors. 
In this test, the performances of the HO scheme are slightly worse
than those of the LLF. The HO-Hyb algorithm gives results in-between
the HO and the LLF-method. Again, for all the schemes, second-order
convergence is observed. 

\begin{figure}[t]
  \centering
  \includegraphics[width=0.49\textwidth]{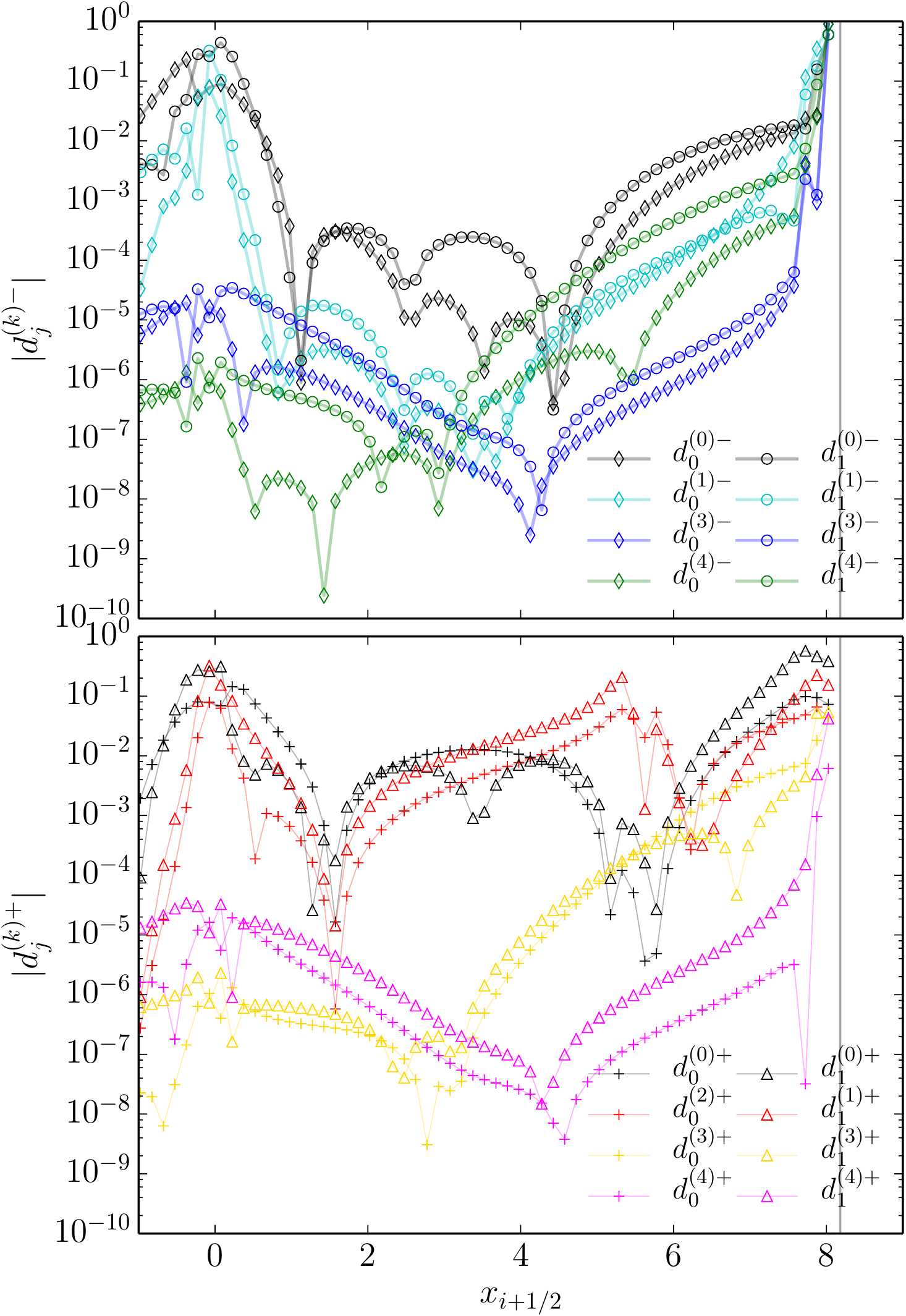}
  \caption{Deviation from optimal WENOZ weights for the $\Gamma=2$ EOS 
    spherical star. 
    The plot shows the absolute difference with respect to optimal weights
    $d^{(k)\pm}_j$ of \eqref{dki} for various 
    projected fluxes $\hat{F}^{(k)\pm}$ and after a
    very short evolution time, $t\sim0.003$.
    Note that, by symmetry, $d^{(2)\pm}_j=d^{(1)\pm}_j$ and
    $d^{(k)\pm}_j(x>0)=d^{(k)\mp}_j(x<0)$. 
    For this test we evolve few Euler timesteps at a
    unigrid resolution of $h=0.15$, $n=128$, and CFL of $0.01$. The
    vertical line marks the star surface.}
   \label{fig:star_wenoweights} 
\end{figure}

\textit{Why the nominal high-order convergence is not achieved in
  these tests?}
By considering the WENO weights for this problem, one finds that, at
least at these resolutions, the weights are not optimal. Hence, the
interpolation of the intercell-fluxes is affected by the lower-order
component, and nominal convergence cannot be
achieved. Figure~\ref{fig:star_wenoweights} shows the absolute values
of the differences 
\be 
\label{dki}
d^{(k)\pm}_j=\omega^{(k)\pm}_j-o_j
\ee
for $j=0,1$ (the third weight $j=2$ carries redundant information and
it is not shown), where 
$\omega^{(k)\pm}_j$ are the weights associated to the reconstruction
of the characteristic flux variables $\hat{F}^{(k)\pm}$. 
The $d_j$'s are computed in the $x$-direction and after a
very short evolution time corresponding to few Euler timesteps. 
For optimal weights, one has $d_j=0$, and the algorithm locally 
achieves the formal fifth-order convergence. If $d_j\gg0$ for some $j$, then
the corresponding sub-stencil has a smaller weight in the
reconstruction and the local convergence order is reduced.
As shown in the figure, the weights significantly deviate from the
optimal values both in the bulk of the star (notably at the center),
and at the surface. A rigorous and complete analysis of the WENO
weights on the star solution in 3D is difficult and beyond the
purpose of this work. However, by inspecting the convergence
properties of the weights in some given direction we observe that some
of the necessary conditions for high-order convergence, namely
$d^{(k)\pm}_j=\O(h^2)$ and $\sum_jd^{(k)\pm}_j=\O(h^6)$
\cite{Henrick2005542}, are not met.

\section{Binary Neutron Star Evolutions}
\label{sec:bns}

\begin{table}[t]
  \centering    
  \caption{BNS quasicircular initial data. Columns: name, EOS, binary
  mass, rest mass, ADM mass, angular momentum, GW frequency. All
  configurations are equal-masses and irrotational.}  
  \begin{tabular}{ccccccccc}        
    \hline
    \hline
    Name & EOS & $M$ & $M_b$ & $M_{\rm ADM}$ & $J_0$ & $M\omega_0$ & \\
    \hline
    SLy135135\_0060  & SLy  & 2.700 & 2.989 & 2.671 & 6.872 & 0.060 \\ 
    SLy135135\_0038  & SLy  & 2.700 & 2.989 & 2.678 & 7.658 & 0.038 \\ 
    MS1b135135\_0038 & MS1b & 2.700 & 2.935 & 2.678 & 7.665 & 0.038 \\ 
    \hline
    \hline
  \end{tabular}
  \label{tab:conf} 
\end{table}

\begin{table*}[t]
  \centering    
  \caption{Grid configurations and summary of BNS runs.
    Columns: 
    BNS configuration, 
    shortname for grid configuration, 
    refinement levels, 
    minimum moving level index, 
    number of points per direction in fixed levels, 
    number of points per direction in moving levels, 
    resolution per direction in the finest level $l=L-1$, 
    number of points in radial direction in spherical patches, 
    resolution per direction in the level $l=0$.}
  \begin{tabular}{c|cccccccc|cc|cc|cc}        
    \hline
    \hline
     BNS & \multicolumn{8}{c|}{Grid} & \multicolumn{6}{c}{Numerical
       Flux Scheme} \\
     & Name & $L$ & $l^{mv}$ & $n$ & $n^{mv}$ & $h_{L-1}$ & $n_r$ &
  $h_0$ & \multicolumn{2}{c}{LLF} & \multicolumn{2}{c}{HO}
  & \multicolumn{2}{c}{HO-Hyb}\\
 & & & & & & & & & MP5 & WENOZ & MP5 & WENOZ & MP5 & WENOZ\\
    \hline
    \multirow{4}{*}{SLy135135\_0060} & Low (L) & 7 & 2 & 128 & 64  & 0.228  & 128 & 14.592 & \check & \check & \check& \check& \check& \check\\
     & Med (M) & 7 & 2 & 192 & 96  & 0.152  & 192 & 9.728& \check & \check & \check& \check& \check& \check  \\
     & Hig (H) & 7 & 2 & 256 & 128 & 0.114  & 256 & 7.296& \check & \check & \check& \check& \check& \check  \\
     & Fin (F) & 7 & 2 & 320 & 160 & 0.0912 & 320 & 5.8368& \check & \check & \check& \check& \check& \check \\
    \hline
    \multirow{4}{*}{SLy135135\_0038} & Low (L) & 7 & 2 & 160 & 64  & 0.228 & - & 14.592& \cross & \check & \cross& \cross& \cross& \check  \\
                                     & Med (M) & 7 & 2 & 256 & 96  & 0.152 & - & 9.728 & \cross & \check & \cross& \cross& \cross& \check   \\
                                     & Hig (H) & 7 & 2 & 320 & 128 & 0.114 & - & 7.296 & \cross & \check & \cross& \cross& \cross& \check  \\
                                     & Fin (F) & 7 & 2 & 400 & 160 & 0.0912 & - & 5.8368 & \cross & \cross & \cross& \cross& \cross& \check  \\     
    \hline
    \multirow{4}{*}{MS1b135135\_0038} & Low (L) & 7 & 2 & 128 & 80  & 0.291   & - & 18.624 & \cross & \check & \cross& \cross& \cross& \check\\
                                      & Med (M) & 7 & 2 & 192 & 120 & 0.21825 & - & 12.416 & \cross & \check & \cross& \cross& \cross& \check \\
                                      & Hig (H) & 7 & 2 & 256 & 160 & 0.1455  & - & 9.312 & \cross & \check & \cross& \cross& \cross& \check\\
                                      & Fin (F) & 7 & 2 & 320 & 200 & 0.1164  & - & 7.4496 & \cross & \cross& \cross& \cross& \cross& \check\\     
    \hline
    \hline
  \end{tabular}
  \label{tab:grid} 
\end{table*}

\subsection{Initial configurations and grid setups}
\label{sec:bns:conf}

Initial data for our simulations are conformally-flat BNS irrotational
configurations in quasicircular orbits computed with the Lorene
library~\cite{Gourgoulhon:2000nn}. They
are characterized by the Arnowitt-Deser-Misner (ADM) mass-energy
$M_\text{ADM}$, the angular momentum $J_0$, baryonic or rest mass
$M_b$, and the mass-rescaled and 
dimensionless GW circular frequency $M\omega_0$. 
The relevant properties are summarized in Tab.~\ref{tab:conf}.
SLy135135\_0060 is described by the SLy EOS and is prepared at about
three orbits to merger. SLy135135\_0038 is a similar configuration at
about 10 orbits to merger. MS1b135135\_0038 is described by the MS1b
EOS, and has same initial frequency of SLy135135\_0038. 

We present a total of 32 new evolutions of these initial
data. SLy135135\_0060 has been evolved with LLF, HO, and HO-Hyb
numerical fluxes, each combined with WENOZ and MP5
reconstructions. For each flux and reconstruction combination, four
different grid resolutions were considered. The grid specifications
for all the runs are reported in Tab.~\ref{tab:grid}. 
The four grid resolutions allow us to perform self-convergence tests 
by choosing differently resolved triplets of data. 
We analyze two different triplets with $n^{mv}=(64,96,128)$ and
$n^{mv}=(96,128,160)$ chosen such that the differences in the three
resolutions are nearly optimal \cite{Bernuzzi:2011aq}. 
SLy135135\_0038 and MS1b135135\_0038 have been evolved only with the
HO-Hyb-WENOZ scheme, but we evolved the same data with the LLF-WENOZ
 in \cite{Bernuzzi:2014owa}. Note also that the runs *\_0038 with
HO-Hyb-WENOZ were evolved without spherical patches, i.e. using a
computationally less expensive setup. Atmosphere parameters are
$f_{atm}=10^{-11}$, $f_{thr}=10^2$, $f_{hyb}=10^2$.

During the development of this work we tested the robustness of our
findings by further varying i) grid setups, ii) the formulation for the metric
equations, and iii) BNS configurations. In particular, we performed
additional convergence 
tests with a grid setup employing neither moving boxes nor spherical
patches (cf.~\cite{Radice:2013xpa}) evolving SLy135135\_0060 with
LLF-WENOZ and HO-WENOZ and using BSSNOK. Other convergence tests were
performed with the $\Gamma$-law EOS configuration previously used
in~\cite{Thierfelder:2011yi,Hilditch:2012fp}. In the following, we do
not discuss these tests but focus on the most detailed and
representative simulations of Tab.~\ref{tab:grid}. All our results are
consistent with the findings reported in the following. 
We expect our findings will carry over in the computation of inspiral-merger
waveforms with the broad range of EOS typically employed for GW
modeling (e.g. \cite{Bernuzzi:2014owa}) and for approximately equal masses BNS.

\subsection{Conserved quantities} 
\label{sec:bns:cons}

\begin{figure}[t]
  \centering    
  \includegraphics[width=0.49\textwidth]{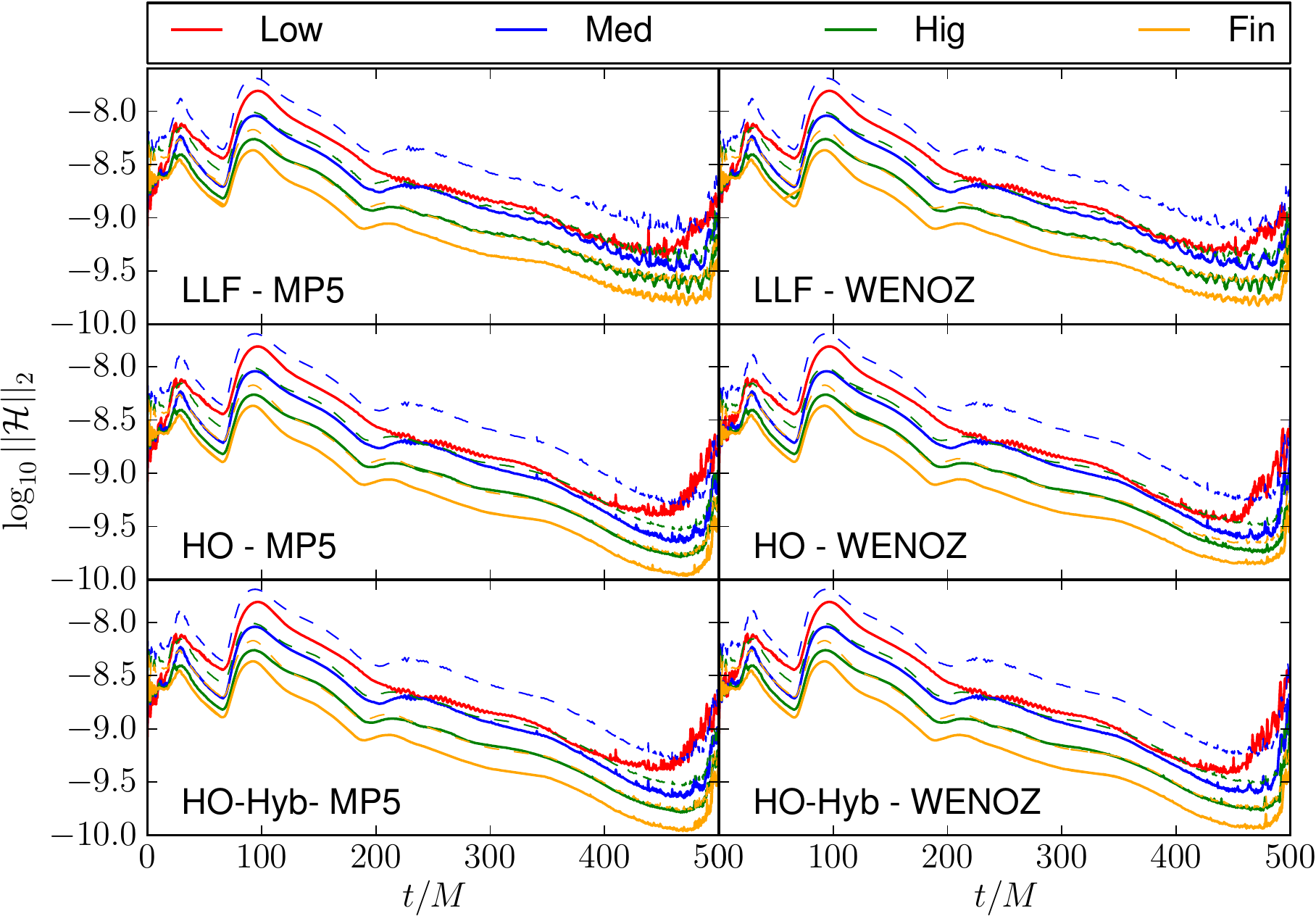}
  \caption{SLy135135\_0060: $L_2$ norm of the Hamiltonian constraint on refinement
  level $l=1$ for different grid resolution and numerical
  schemes. Dashed lines show results scaled to second-order.}
    \label{fig:bns_cons_Ham}
\end{figure}

\begin{figure}[t]
  \centering    
  \includegraphics[width=0.49\textwidth]{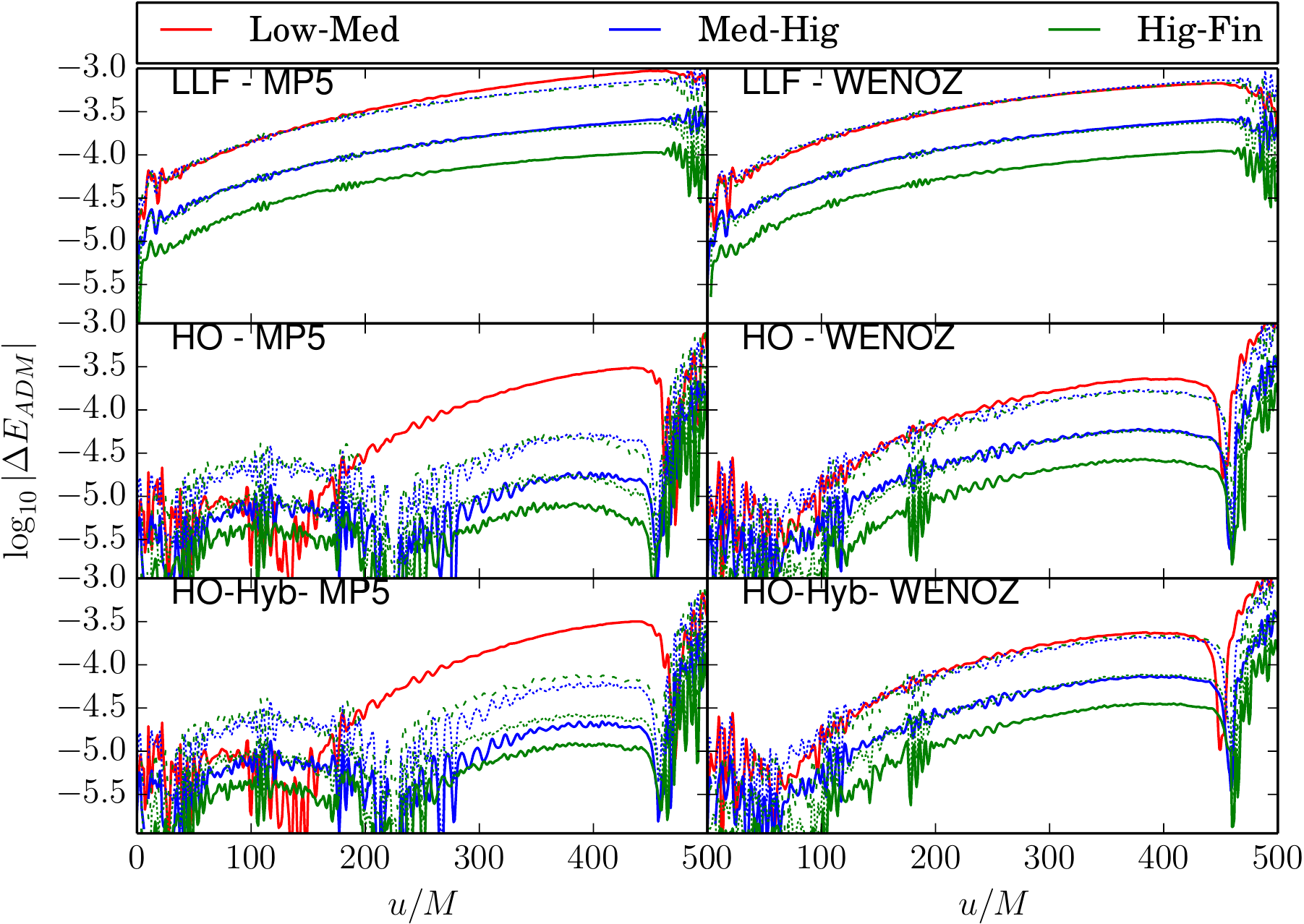}
  \caption{SLy135135\_0060: ADM energy Eq.~\eqref{MADM} extracted at coordinate sphere $r=450$ for
    different grid resolution and numerical schemes. Dashed lines show
    results scaled to second-order.} 
 \label{fig:bns_cons_MADM}
\end{figure}

\begin{figure}[t]
  \centering    
  \includegraphics[width=0.49\textwidth]{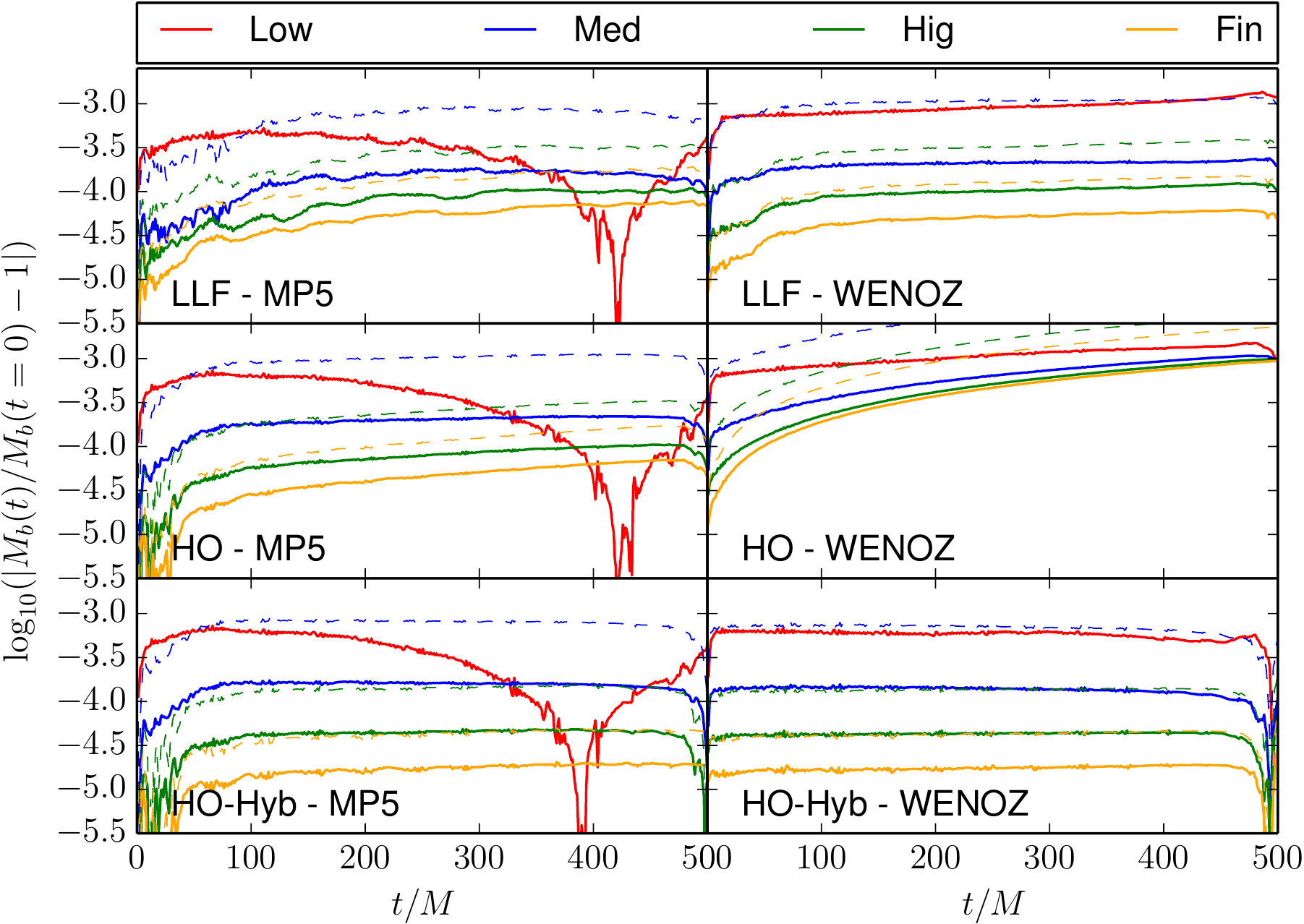}
  \caption{SLy135135\_0060: rest mass conservation $\log_{10} \left(
    \left|M_{b}(t)/M_{b}(t=0)\right|\right)$  
  and its convergence on refinement level
  $l=5$ for different grid resolution and numerical
  schemes. Dashed lines show results scaled to fourth-order; which is the
  order of the restriction operator (see text).}
 \label{fig:bns_cons_Mb}
\end{figure}

\begin{figure}[t]
  \centering    
  \includegraphics[width=0.49\textwidth]{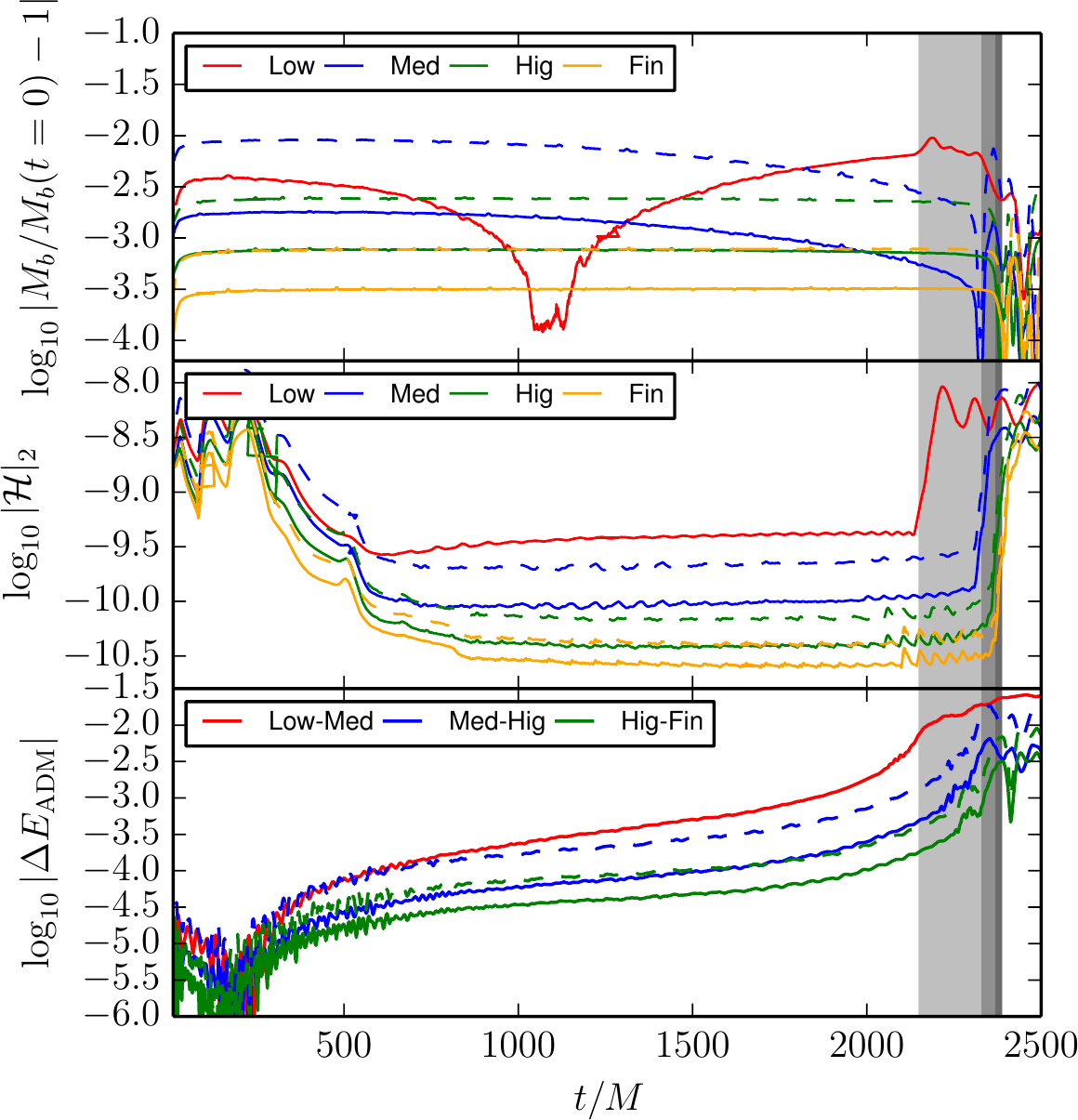}
  \caption{Conserved quantities for SLy135135\_0038 with the
    HO-Hyb-WENOZ scheme. 
  Top panel: conservation of the baryonic mass measures on level $l=4$. 
             Dashed lines correspond to an assumed fourth order
             convergence, cf.~Fig.~\ref{fig:bns_cons_Mb}. 
  Middle panel: $L_2$-volume norm of the Hamiltonian constraint
  measured on level $l=1$.  
             Dashed lines correspond to an assumed second-order convergence, cf.~Fig.~\ref{fig:bns_cons_Ham}.
  Bottom panel: Self-convergence of the ADM-energy, where we have not corrected the energy by the emitted 
             GW energy. The energies are extracted at $r=1000M_\odot$. 
             Dashed lines correspond to an assumed second-order
             convergence, cf.~Fig.~\ref{fig:bns_cons_MADM}.}
  \label{fig:bns_diagnostics_sly0038}
\end{figure}

Conserved quantities are fundamental diagnostics for evaluating the
performance of a numerical scheme. Before discussing the waveform
accuracy we discuss the convergence properties of these quantities. During
the BNS evolution we choose to monitor the following three quantities. 

\textit{(1)} The $L_2$ norm of the Hamiltonian constraint,
$||\mathcal{H}||_2$, see e.g.~\cite{Hilditch:2012fp} for the
definition of $\mathcal{H}$. In the continuum limit
$\mathcal{H}\to0$, any numerical solution must show convergence to
zero in order to be consistent with Einstein equations. 

\textit{(2)} The ADM mass of the spacetime. In asymptotically flat spacetimes,
  one can define the conserved ADM mass as $M_{\rm
  ADM}=\lim_{r\to\infty} E_{\rm ADM}(r)$, where 
\be
\label{MADM}
E_{\rm ADM}(r)=\int_{S_r} d s_l
\sqrt{\gamma}\,\gamma_{ij} \gamma^{kl}\left( \gamma_{ik,j} - \gamma_{ij,k}\right) 
\ee
is computed on a coordinate spheres $S_r$ of radius $r$ in the wave
zone, and $\gamma_{ij}$ is the spatial metric. $E_{\rm ADM}(r)$ is
expected to deviate from $M_{\rm ADM}$ due to the 
gravitational energy radiated away from the sphere, 
$E_\text{GW}$. 
Conservation at large finite $r$ is
expected for the quantity $M_{\rm ADM}^{(r)}=E_{\rm ADM}(r)+E_\text{GW}(r)\approx
M_{\rm ADM}$.  

\textit{(3)} The rest mass of the matter, 
\be
 \label{Mb}
 M_{b} = \int d^3 x\sqrt{\gamma}D
\ee
whose conservation follows from the first component of \eqref{hydro},
see e.g.~\cite{Font:2007zz,Dietrich:2015iva}. In case \eqref{hydro} are solved on a
single grid, $M_b$ is preserved at round-off error; violations at the
truncation error level are, however, generically expected in cases in
which adaptive mesh refinement is employed~\cite{Dietrich:2015iva}. 

We first discuss the conservation and convergence of these quantities in 
the SLy135135\_0060 evolutions for different numerical schemes and
resolutions, then consider the same quantities for the longer runs.

\subsubsection{SLy135135\_0060: Hamiltonian constraint}

The $L_2$ norm of the Hamiltonian constraint computed on level $l=1$ is shown in
Fig.~\ref{fig:bns_cons_Ham}. During the inspiral-merger the norm is
decreasing because of the constraint propagation and damping  
properties of the Z4c evolution scheme~\cite{Weyhausen:2011cg}. The
violation of the constraint is of the order $\sim10^{-9}$ and shows 2nd
order convergence to round off at increasing grid resolution (cf. dashed
lines). Note that constraint violations  
stay below the initial value. We observe no significant difference between 
the different reconstruction and flux computation routines. 

\subsubsection{SLy135135\_0060: ADM energy}

The convergence of $E_{\rm ADM}$ at increasing grid resolutions
is shown in Fig.~\ref{fig:bns_cons_MADM}. WENO-based schemes achieve 
clean second-order convergence up to the moment of merger. The use of
high-order schemes for the numerical fluxes results in smaller absolute
differences between one resolution and the other, thus indicating a
better accuracy. By contrast, the combination of high-order schemes
and MP5 reconstruction performs worse. Although, MP5 gives smaller
differences for high resolutions, one does not see a clear convergence
order, and the Low (L) grid is not in the convergent regime. 

Overall, the relative accuracy of $E_{\rm ADM}$ for the grid F is of
the order of $10^{-3}-10^{-4}$. That accuracy is signifcantly smaller
than the GW energy emitted $E_{\rm GW}(r)$ only towards the merger moment, where $E_{\rm
  GW}(r)$ reaches few percent~\cite{Bernuzzi:2015opx}. As a consequence, the
convergence properties of the complete quantity $M_{\rm ADM}^{(r)}$ are
slightly worse and, as with high-order methods, it remains difficult
to properly resolve $M_{\rm ADM}^{(r)}$. \\

Both $\mathcal{H}$ and $E_{\rm ADM}$ data show that the truncation
error scales at second-order at the considered resolutions. Motivated by
the analysis of Sec.~\ref{sec:ss}, we have inspected the WENO weights
also for the binary case. The quantities $d^{(k)\pm}_i$ for the stars
in the binary are qualitatively similar, at early times, to the single star
case. The HO schemes fail to achieve the nominal
high-order convergence both in the bulk of the matter and at the 
surface. We conclude that high-order convergence cannot be expected at
these resolutions. 

\subsubsection{SLy135135\_0060: Rest Mass}

Let us discuss rest mass conservation. In neutron star simulations the 
conservation of rest mass is affected by three dominant errors:
(i) the artificial atmosphere, (ii) the restriction operation, and 
(iii) refinement boundaries. 
Because of our particular grid choice no significant rest mass crosses
the finest 
refinement level during the inspiral and merger. 
When the two boxes of the finest refinement level touch each other, 
BAM regrids and a larger box covering the 
strong-field region is created. Thus, we expect that artificial atmosphere
and restriction are the main source of uncertainty in Eq.~\eqref{Mb}.
Of course, the situation is different in the subsequent postmerger and collapse
phase in which the rest mass spread all over the grid and the
particular choice of slicing leads to an apparent loss of rest mass
when the singularity forms \cite{Thierfelder:2010dv,Dietrich:2014wja}. 
In this paper, we focus on the inspiral-merger and do not discuss these issues.
The artificial atmosphere treatment typically introduces a
$\mathcal{O}(h^2)$ error in $M_b$ that is characterized by mass losses, see
\cite{Dietrich:2015iva}. Note that an error localized around
the star surface can become negligible for the volume integral
\eqref{Mb} at sufficiently high resolutions. 
The restriction operation and regridding from the finer levels to coarser
ones are performed in this work with a fourth-order WENO algorithm. Thus, a 
$\mathcal{O}(h^4)$ error is expected at best. 

The rest mass conservation is monitored by the 
relative error $\left|M_{b}(t)/M_{b} (0)\right|$,
where $M_{b}(t=0)$ is computed after interpolation of the initial data
on the BAM grid and after restriction of the data from finer to
coarser levels. The evolution of the error is shown in
Fig.~\ref{fig:bns_cons_Mb}.
For any reconstruction choice, the HO schemes
perform significantly worse than the LLF or HO-Hyb schemes. 
This is most likely due to a lack of accuracy in the resolution of 
the low-density flow, caused by the particularly large values of 
$f_{atm}$ necessary to stabilize these
simulations. This effect is observed also in the single star
evolutions of Sec.~\ref{sec:ss}, although we have not discussed this 
explicitly there.
At the lowest resolution the use of MP5 reconstruction produces
serious violations of $M_b$ (red lines, right panels). 
A convergence order $\sim 4$ is observed in the LLF-WENOZ runs at
early times; for $t\gtrsim100M$ the convergence of the error is slower
and dominated by the $\mathcal{O}(h^2)$ component. 
The best conservation is obtained with the HO-Hyb flux scheme 
employing the WENOZ reconstruction. In these simulations the dominant
error is given by the truncation error $\O(h^4)$ of the restriction operator
over the whole simulated time.

\subsubsection{SLy135135\_0038 and MS1b135135\_0038}

The analysis presented for the SLy135135\_0060 evolutions
holds qualitatively also for the longer and more challenging 10 orbits
simulations, although we find that for such long simulations higher resolutions 
are needed to obtain the same accuracy. In particular, overconvergence
is observed in one triplet and it is due to the lowest resolution simulations.

Conserved quantities are shown in
Fig.~\ref{fig:bns_diagnostics_sly0038} for SLy135135\_0038.
The upper panel shows the rest mass conservation. 
In contrast to SLy135135\_0060, the two lowest resolutions do not show a clear convergence order. 
The error grows up to 1\% on level $l=4$. 
For the two highest resolutions fourth order convergence, which is again caused by the restriction operation, is observed. 
For the highest resolution the error stays during the entire inspiral below the 0.05\% level within level $l=4$. 
The error even decreases significantly when computing the mass on level $l=5$ and becomes smaller than $10^{-5}$ 
for the highest resolution. This shows that, for the considered levels, the restriction operation is the 
dominant error contribution for the two highest resolutions and supports our argumentation for 
SLy135135\_0060.

The middle panel of Fig.~\ref{fig:bns_diagnostics_sly0038} presents the evolution of the $L_2$-volume norm of the Hamiltonian 
constraint. As for SLy135135\_0060, the constraints stay below the initial value during the entire inspiral up to the moment of merger. 
second-order convergence is recovered for the two highest resolutions, cf.~dashed lines. 

The bottom panel of Fig.~\ref{fig:bns_diagnostics_sly0038} shows a self-convergence test of the ADM-energy extracted at $r=1000M_\odot$. 
The dashed lines correspond to second-order convergence, which is obtained for the two highest resolutions, but not for the lower resolution. 
The triplet (L,M,H) shows overconvergence due to the
Low (L) resolution simulation.

\subsection{Waveform convergence}
\label{sec:conv}

\begin{figure*}[t]
  \centering    
  \includegraphics[width=\textwidth]{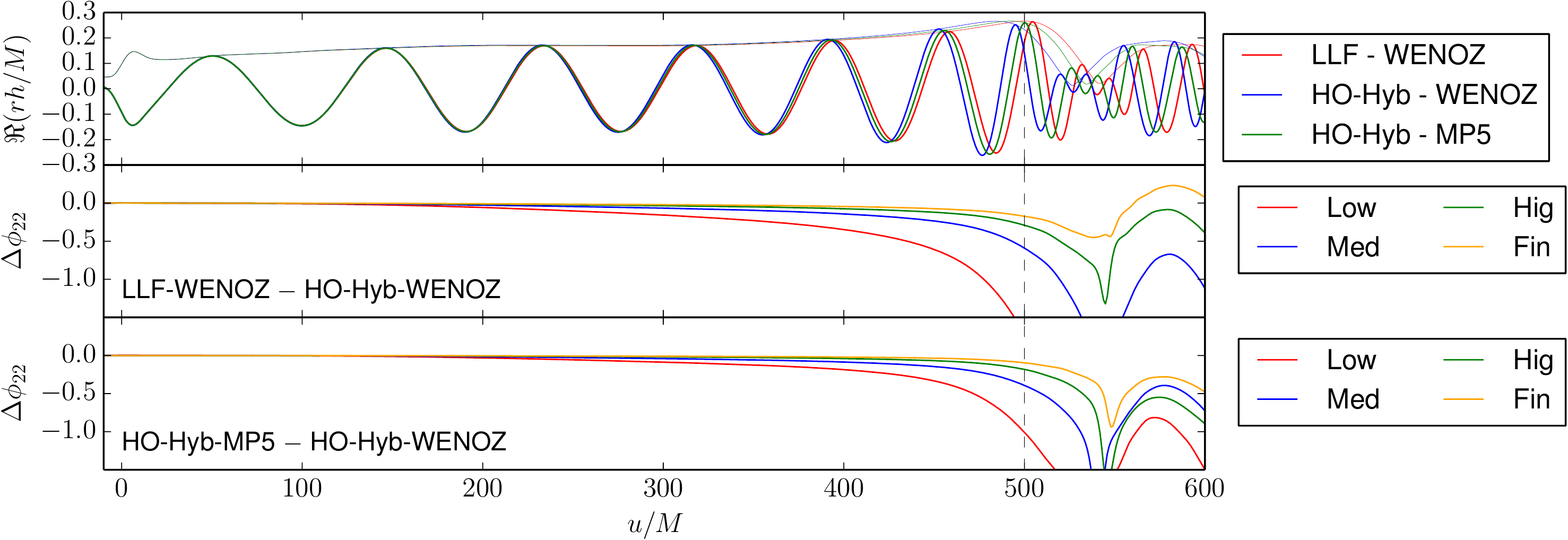}\\
  \caption{Top: $\ell=m=2$ waveforms for SLy135135\_0060 from runs
    with grid Low and varying the numerical scheme. 
    The vertical line marks the moment of merger for the LLF simulation.
    Middle: Phase differences between HO-Hyb and LLF simulations employing WENOZ
    reconstruction and with the same grid resolutions.
    Bottom: Phase differences between HO-Hyb simulations employing
    WENOZ and MP5 reconstruction with the same grid resolution.} 
  \label{fig:bns_phidiff_sly}
\end{figure*}

\begin{figure*}[t]
  \centering    
 \includegraphics[width=0.9\textwidth]{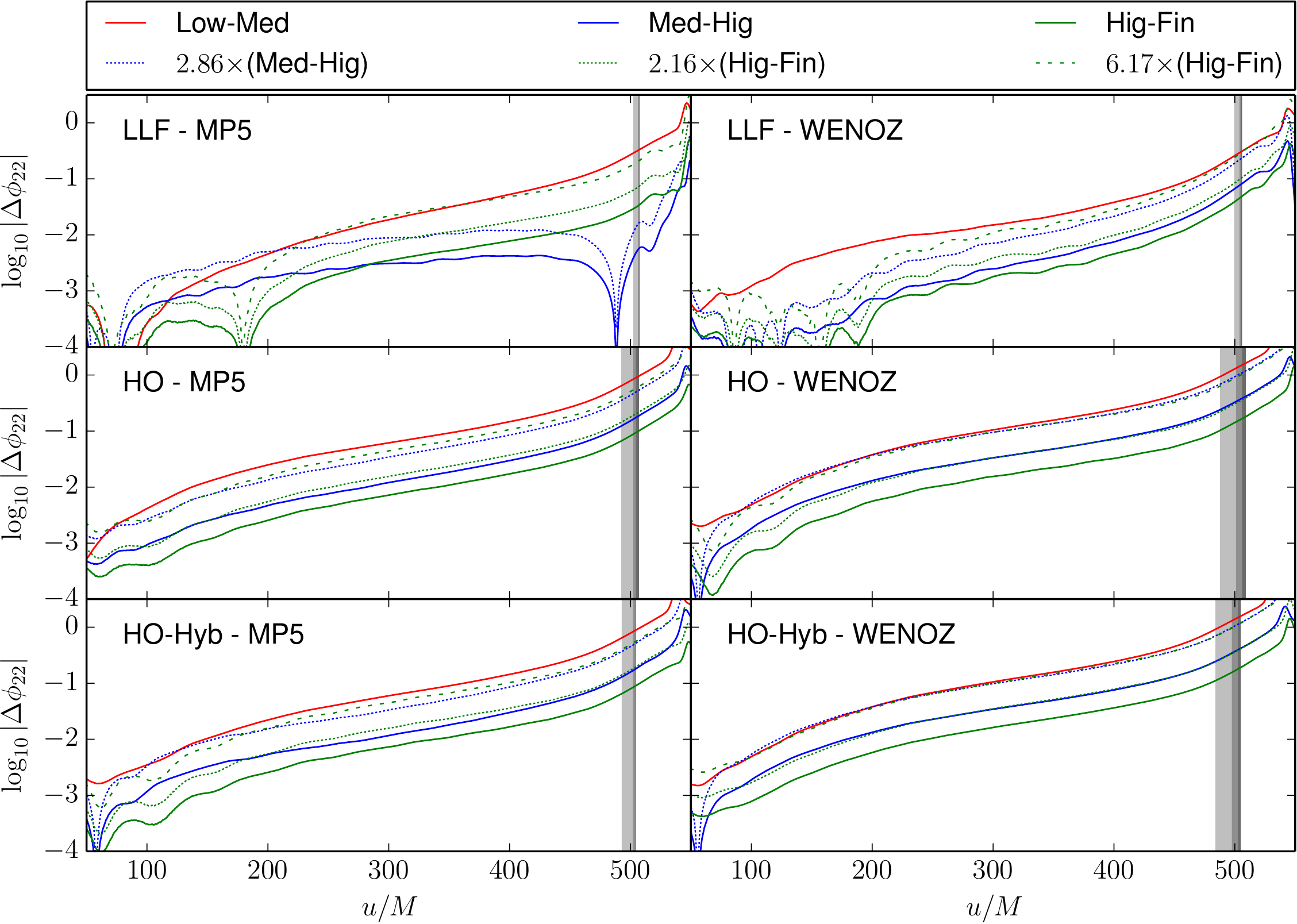}  
  \caption{Convergence of SLy135135\_006. 
    The different panels show the phase differences in log scale for
    different reconstructions: MP5 (left) and WENOZ (right). 
    The algorithms from top to bottom are: LLF, HO, HO-Hyb.
    The vertical shaded regions represent the moment of merger for different resolutions: 
    light gray for $[u_{\rm mrg}^{\rm L},u_{\rm mrg}^{\rm M}]$, gray for 
     $[u_{\rm mrg}^{\rm M},u_{\rm mrg}^{\rm H}]$ and dark gray for 
      $[u_{\rm mrg}^{\rm H},u_{\rm mrg}^{\rm F}]$.
    To show convergence we rescale the phase differences 
    assuming second-order convergence, cf.~dashed and 
    dot-dashed lines.}
  \label{fig:bns_phiconv_sly}
\end{figure*}

In this section we discuss the convergence of the GWs. We focus on the
GW phase that is the main quantity of interest for GW modeling. A major
goal for NR simulation is to control phase uncertainty and assign precise
error bars to this quantity. 

The emitted GWs are calculated by computing the Weyl $\Psi_4$ scalar on
coordinate spheres of radius $r$ in the wave zone. As customary, we
work with the spin weighted spherical harmonics projections (independent of the
viewing angle), reconstruct the GW multipolar modes $h_{\ell m}$ from
$\Psi_4$ projections by solving $\ddot{h}_{\ell m} = \psi_{\ell m}$ in
the frequency 
domain~\cite{Reisswig:2010di}, and focus on the dominant mode
$\ell=m=2$. Amplitude and phase are defined as 
\be
r\, h_{22} = A_{22} \ e^{- i \phi_{22}} \ .
\ee
Waveforms are shown against the retarded time 
\be
u = t-r_* = t-r-2M\log\left(\frac{r}{2M}-1\right) \ .
\ee
The {\it moment of
merger} $u_\text{mrg}$ is defined as the time of the first peak of $A_{22}$, and conventionally
marks the end of the inspiral \cite{Bernuzzi:2014kca}. 

\subsubsection{SLy135135\_0060}

Figure~\ref{fig:bns_phidiff_sly} show the real part of the
SLy135135\_0060 waveform at the lowest resolution for different
numerical schemes (top panel) and the phase difference obtained using
different numerical schemes and the same resolutions. The vertical line
marks $u_\text{mrg}\sim500M$ for the data of
grid Low. 
The phase differences of waveforms computed with different numerical
scheme decrease as resolution increases, indicating that all 
schemes converge to the same continuum, physical solution. 
At $u_\text{mrg}$ the phase differences are largest. The differences
between HO-Hyb and LLF (both with WENOZ 
reconstruction) at the lowest
resolutions are $\Delta\phi\sim-2$~rad, and reduce to 
$\Delta\phi\sim-0.15$~rad at the highest resolution. 
The differences between HO-Hyb-MP5 and HO-Hyb-WENOZ at the lowest
resolutions are $\Delta\phi\sim-1$~rad, and reduce to 
$\Delta\phi\sim-0.1$~rad at the highest resolution.
The difference between HO-WENOZ and HO-Hyb-WENOZ at the lowest
resolution is of the order $\Delta\phi\sim-0.5$~rad, and it reduces to
$\Delta\phi\sim-0.25$~rad  
at the highest resolution (not shown in the figure). 

\begin{figure}[t]
  \centering    
  \includegraphics[width=0.49\textwidth]{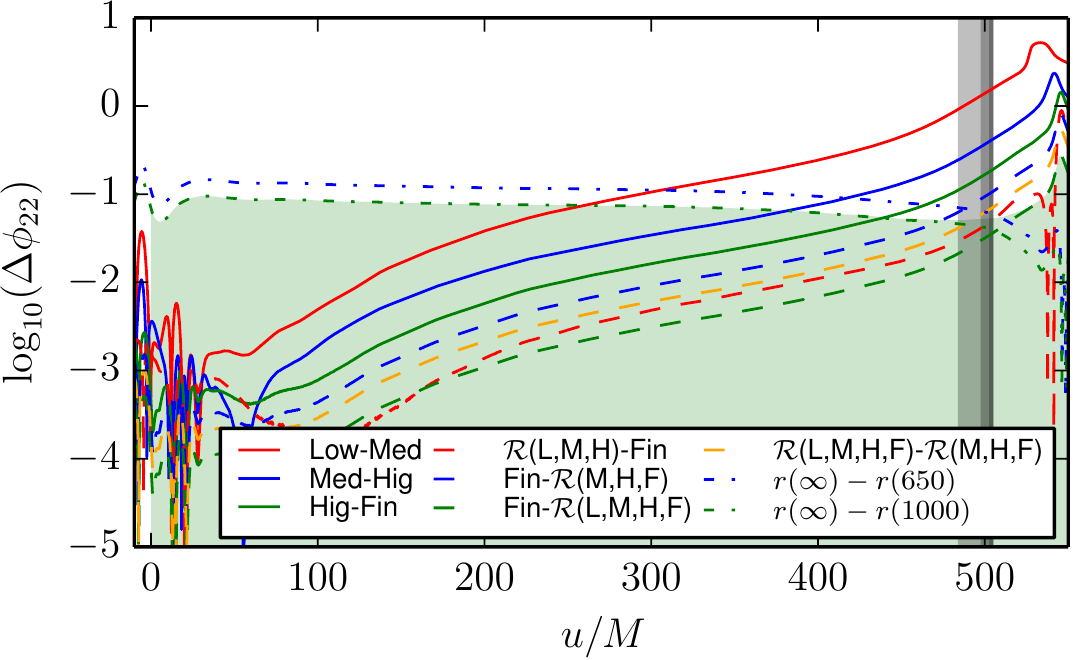}
  \caption{Phase error budget of SLy135135\_0060 with the HO-Hyb-WENOZ
    scheme. Solid lines are 
    phase differences between runs at different resolutions; dashed
    lines are differences with the Richardson extrapolated data;
    dashed dotted lines are differences between phase extracted at
    finite radii and the radius-extrapolated one.}
  \label{fig:bns_errbudget_sly}
\end{figure}

A three-level self-convergence study for SLy135135\_0060 is shown in
Fig.~\ref{fig:bns_phiconv_sly} for all the numerical schemes
considered. As mentioned above, performing the analysis with four  
instead of three different resolutions allows one to robustly assess
convergence. The latter is, in fact, evaluated for two triplets. 
All our convergence plots show both the differences between the data
(solid lines) and those scaled assuming second-order convergence
(dotted and dashed lines). Note that, in all our simulations, the
phase is overestimated (faster) at low resolutions as a result of
numerical dissipation \cite{Bernuzzi:2012ci}.

Independently on the reconstruction method, the LLF scheme shows larger
oscillations and a more complicated behavior than the corresponding HO
and HO-Hyb simulations. The convergence properties of LLF-MP5 data are not
systematic and a zero crossing in $\Delta \phi$ is visible before
$u_\text{mrg}$. These LLF-MP5 data are not in convergence regime and
are unreliable for producing an error budget. The LLF-WENOZ data
improve on this situation. With the LLF-WENOZ setup we observe a
convergence order slightly higher than second-order in the triplet
(L,M,H), and slightly below second-order in the triplet
(M,H,F). That indicates that the resolution $n=64$ is too low and
should be discarded; only (M,H,F) should be considered. Still,
it is difficult to obtain a precise error estimate using only the
(M,H,F) triplet because one does not
know how robust is the convergence result at higher
resolutions~\cite{Bernuzzi:2011aq}. Another resolution is desirable;
in the absence of that, the differences between the two
highest resolutions can be taken as an error
estimate~\cite{Bernuzzi:2014owa}.

The HO and HO-Hyb schemes employing the MP5 reconstruction show a
convergence order of $\sim3$ at low resolutions -(L,M,H) triplets-, but
a lower convergence order of $\sim2$ (or slightly below) at higher resolutions, 
(M,H,F,) triplet. The MP5 high-order fluxes improve significantly
over the LLF ones. However, without a fifth resolution, it is 
impossible to make a robust claim about convergence.

Contrary to the other setups, the HO and HO-Hyb WENOZ schemes show a
very robust convergence behavior. We observe clean second-order
convergence for all four resolutions. This convergence order is also 
consistent with what we observe in single star evolution and in the
conserved quantities discussed above. From the analysis of the phasing
solely one cannot clearly point out differences between the HO and
HO-Hyb schemes. However, combining all the results  (cf. Fig.~\ref{fig:star_l1_fgr} and
Fig.~\ref{fig:bns_cons_Mb}), we conclude that the HO-Hyb-WENOZ scheme
performs better among those we tested and at the resolutions explored here.

Let us mention that the plot also shows that the differences between
different resolutions are smaller for LLF simulations than the HO or
HO-Hyb ones. This fact can be interpreted as the LLF having smaller absolute
errors, i.e.~begin more accurate for lower resolutions, although the
interpretation is not straightforward in the absence of clear convergence.

\subsubsection{SLy135135\_0038 and MS1b135135\_0038}

\begin{figure}[t]
  \centering    
  \includegraphics[width=0.49\textwidth]{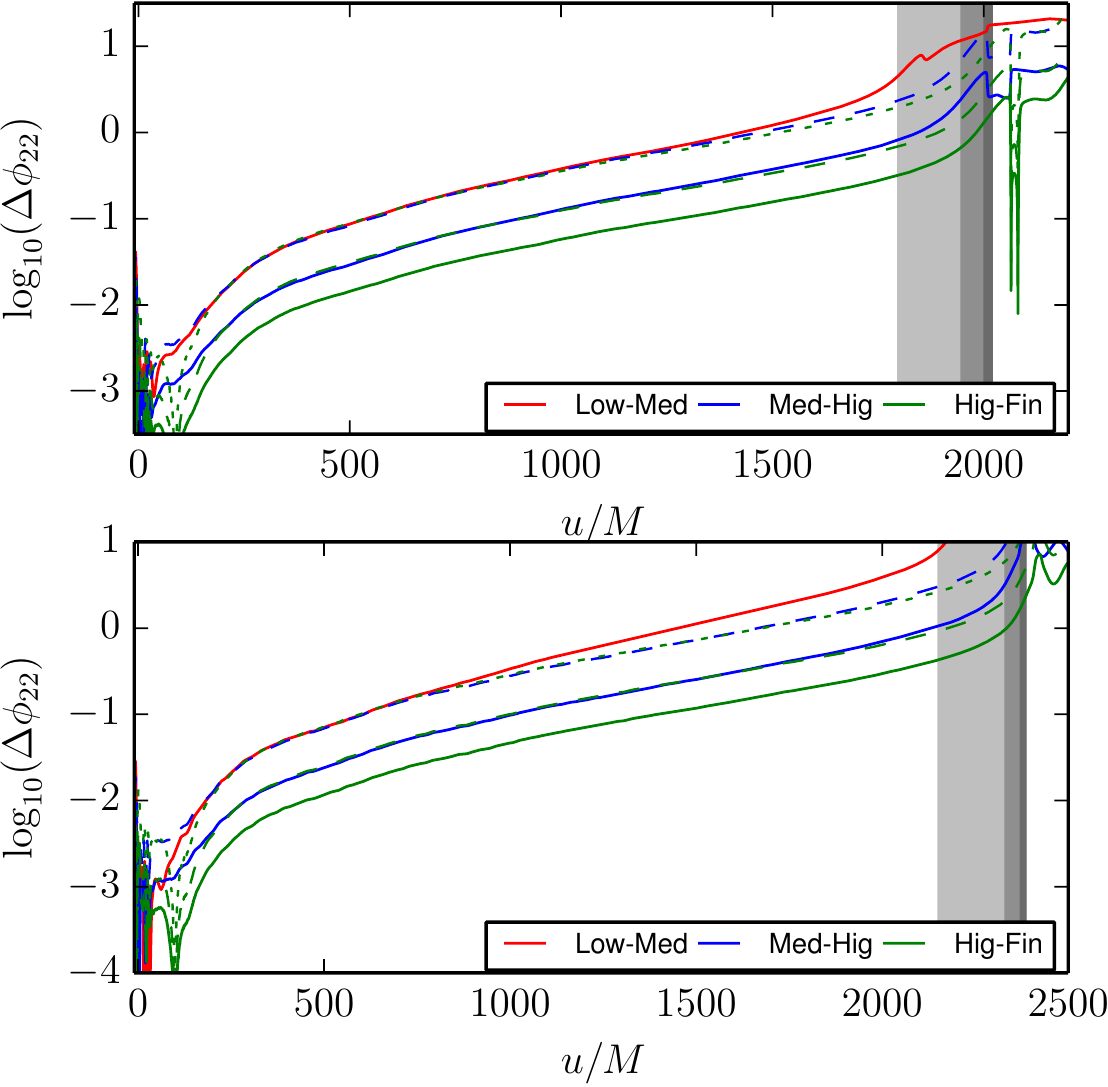}
  \caption{Phase convergence of MS1b135135\_0038 (upper panel) and
    SLy135135\_0038 (lower panel) with the HO-Hyb-WENOZ
    scheme.}
  \label{fig:bns_phiconv_slyms1b}
\end{figure}

Self-convergence studies for the phase evolution of SLy135135\_0038
and MS1b135135\_0038 are shown in Fig.~\ref{fig:bns_phiconv_slyms1b} for
the HO-Hyb-WENOZ scheme. Also in these long runs numerical dissipation
artificially accelerates the phase evolution. 
The convergence behaviour is similar to the
SLy135135\_0060 case, although in these long runs the lowest
resolution $h\sim0.23$ ($n^{mv}=64$) does \textit{not} give convergence results
for $u\gtrsim1000M$ (SLy) and $u\gtrsim1500M$ (MS1b). At late times,
we observe instead overconvergence due to the Low-grid simulation.
As a consequence, the lowest resolution runs should not be used for producing
the error budget. The phase differences accumulated to $u_\text{mrg}$
between H and F-runs are about $\Delta\phi\sim1$ rad. They are about a factor 10
larger than in the shorter runs.

\section{Radius extrapolation}
\label{sec:rad-extrap}

\begin{figure}[t]
  \centering    
  \includegraphics[width=0.49\textwidth]{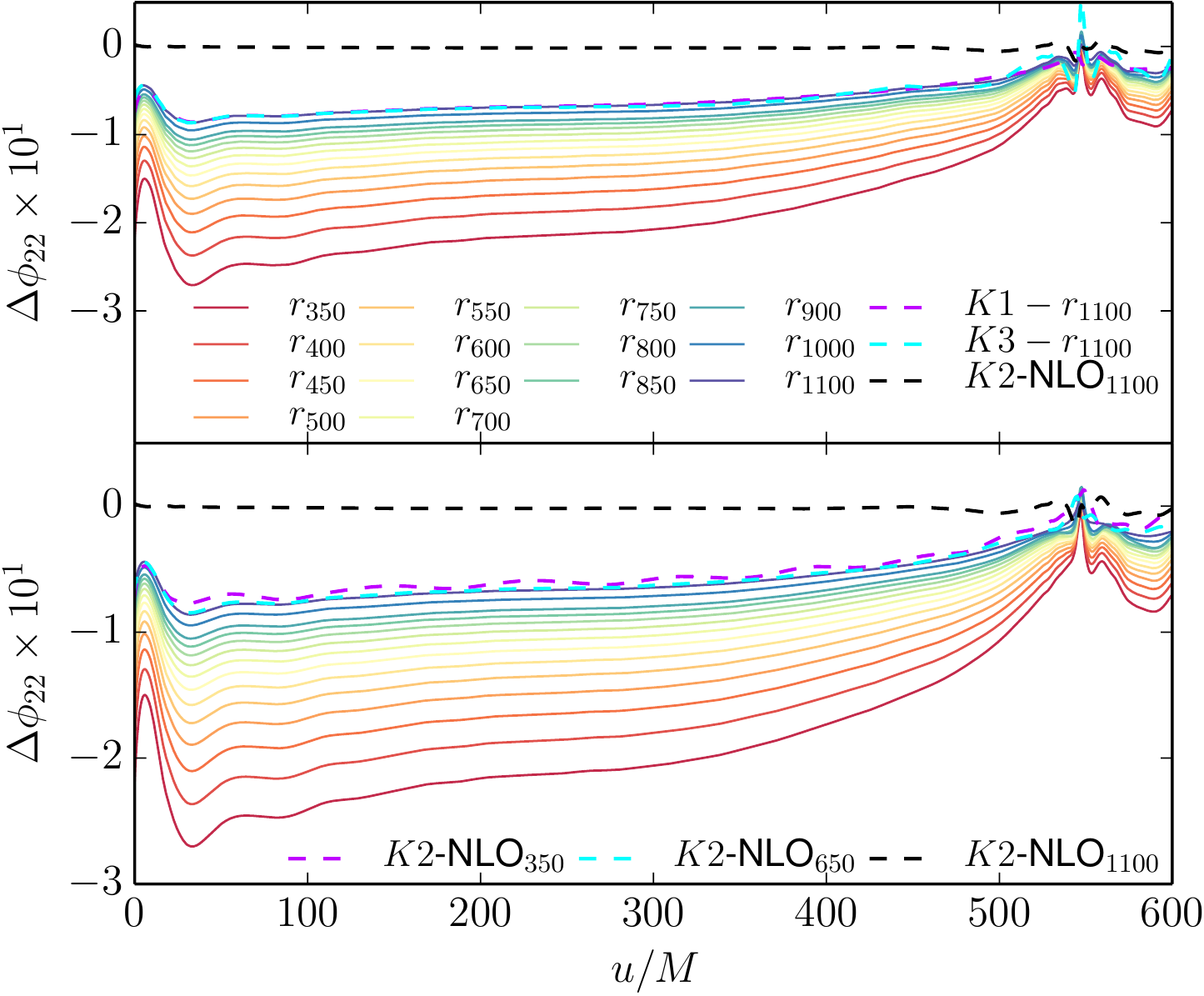}
  \caption{SLy135135\_0060 finite radius extraction uncertainty (HO-Hyb-WENOZ
    scheme).
    Top panel shows the phase difference for different extraction
    radii computed to the   
    polynomial extrapolation with $K=2$ (solid lines). 
    The dashed lines show the difference the extrapolated waves with $K=1,3$ and the last radius $r_N=1100$
    as well as the difference between the $K=2$ and NLO extrapolation. 
    Bottom shows the phase difference between the NLO extrapolated value and finite radii extracted data (solid lines).
    Dashed lines show the difference between the NLO extrapolation for different radii 
    and the $K=2$ extrapolated one.}
  \label{fig:bns_errradius_sly}
\end{figure}

The waveform error budget must take into account also the effect of
finite-radius extraction. As we see, the latter uncertainty is
non-negligible and comparable to the truncation error. It can be evaluated
by extrapolating the waveform in radius
e.g.~\cite{Boyle:2007ft,Bernuzzi:2011aq,Lousto:2010qx}. Notably, the
finite-extraction-radius uncertainty on the phase has opposite sign than
the truncation error uncertainty: the phase evolution is faster for
waveforms extracted at larger radii.  

Two extrapolation methods are considered here. In the first, the waveform
is evaluated at 
different radii $r_j$ with $j=0...N$ and phase and amplitude are extrapolated 
using a polymomial of order $K<N$, 
\be
\label{poly_extrar}
f(u; r_j) = f_0(u) + \sum^K_{k=1} f_k(u) r^{-k}_j \ .
\ee
In the second method, one considers the next-to-leading-order (NLO)
behaviour in $r$ of the $\Psi_4$ multipoles~\cite{Lousto:2010qx}, 
\be
\label{psilm_extrar}
r\,\psi_{\ell m} = r\,\ddot{h}_{\ell m} +
\frac{\ell(\ell+1)}{2r}\; r\,\dot{h}_{\ell m} + \O(r^{-2}) \ , 
\ee
and obtains an extrapolation formula for $r\,\ddot{h}_{\ell m}$ that uses
$\psi_{\ell m}$ extracted at a given radius~\footnote{Note that with a
  slight abuse of notation we indicate with $r\,\psi_{\ell m}$ and $r\,\ddot{h}_{\ell m}$ the
  leading-order asymptotic behaviour.}. The extrapolated $r\,h_{\ell
  m}$ is then reconstructed from $r\,\ddot{h}_{\ell m}$.

Figure~\ref{fig:bns_errradius_sly} compares the differences between
the extrapolated quantities and the finite-radii ones for
SLy135135\_0060 (again, we work with the best data
HO-Hyb-WENOZ). The polynomial extrapolation \eqref{poly_extrar}
requires a choice for $K$, but the phase extrapolation is rather
insensitive on the choices $K=1,2,3$. Considering the 
difference between the extrapolated and the last radius ($r_N=1100$),
one can assign an uncertainty to the extrapolate phase. 
The uncertainty is larger at early times (lower GW frequencies),
$\delta\phi\sim0.1$, and monotonically decreases toward the moment of merger to
$\delta\phi\sim0.05$ 
The fact that the finite-radius uncertainty is larger at lower
frequencies is clear from Eq.~\eqref{psilm_extrar}: taking the double
integral of both sides in order to obtain the metric waveform, and
considering a signal at frequency $\omega$, the second term
$\propto1/r$ gets a factor $1/\omega$ from the integral of the metric
waveform \cite{Lousto:2010qx}. 
 
Figure~\ref{fig:bns_errradius_sly} also compares the polynomial
extrapolation with the second method outlined above. We use 
\eqref{psilm_extrar} for radii $r=350$, $650$ and $r=r_N$.
We find the results are compatible with those of the polynomial
extrapolation. The difference between the extrapolated phases with 
\eqref{poly_extrar} and \eqref{psilm_extrar} is significantly smaller
than the finite radius extraction uncertainty assigned above.
Moreover, the formula \eqref{psilm_extrar} is rather robust for the phase
extrapolation; the use of $r=350$ is already comparable with 
the polynomially extrapolated phase.

We find similar results for the extrapolation of the amplitude (not
shown in the figure). The only difference is that the amplitude
radius-extrapolation is slightly more sensitive to $K$ and stability
can be achieved for $K=2,3$. The uncertainty of the extrapolated
amplitude ($K=2$) is about $0.2\%$ and has smaller variation in time
than the phase.

\section{Waveform Error Budget}
\label{sec:err}

In this section we discuss the error budget on the waveforms generated
with the high-order WENO algorithm. We work with the best data
HO-Hyb-WENOZ, for which we have a clear convergence assessment, and
include the finite-extraction-radius uncertainties discussed above. 

For any finite-differencing algorithm, the value of a quantity $f^{(h)}$
computed at resolution $h$ can be written as 
\be
f^{(h)} = f^{(e)} + \sum^{\infty}_{i=p} A_i h^i \ ,
\ee
where $f^{(e)}$ is the exact, continuum value obtained for
$h\to0$, and $p$ the convergence order. The quantities $A_i$ are not
necessarily small and 
might depend on the number of floating point operations. 
Although it is not possible to compute $f^{(e)}$, one can use the set
of data at different resolutions to improve the simulation results.
A way to proceed is to consider the Richardson extrapolation
method. The latter is a simple algorithm that, given (i) a dataset
$(f^{(h)})$ at different finite resolutions and (ii) an accurate
measure of the convergence order $p$, allows one
to generate a better approximation to $f^{(e)}$~\cite{Richardson:1911}.
The Richardson extrapolated value is indicated
as $\Rich[(f^{(h)})]$. This method has already been 
used for estimating waveform errorbars in
e.g.~\cite{Bernuzzi:2011aq,Hinder:2013oqa,Radice:2015nva}. 
Here we are particularly interested in the robustness/stability of the
extrapolation using different datasets. An alternative resolution
extrapolation method proposed in the literature is studied in
Appendix~\ref{app:extr2}.  

In the following, we consider the Richardson extrapolation of different time series:
$\Rich[(L,M,H)]$, $\Rich[(M,H,F)]$, and
$\Rich[(L,M,H,F)]$, where L stands for grid Low, M for grid Med, and so
on. We investigate whether the different $\Rich[.]$ are consistent.
The extrapolated time series can then be used as best data,
i.e.~an improved approximation of $f^{(e)}$. A measure of the 
uncertainty to be assigned to this best approximation is the
difference with the highest resolution (F) or with another
extrapolation. 

\subsection{SLy135135\_0060}

Figure~\ref{fig:bns_errbudget_sly} shows the waveform phase
differences for SLy135135\_0060 (solid lines) and the differences
between the extrapolated time series and data at grid F (dashed lines). 
The extrapolation $\Rich[(L,M,H)]$ is rather effective in improving
the approximation: it is very close to the phase computed with grid
$F$, although it overestimates the latter.
The positive phase difference $\Delta\phi(H-\Rich[(L,M,H)])$  
is larger than $\Delta\phi(F-\Rich[(L,M,H,F)])$, suggesting a
convergent behaviour. $\Rich[(L,M,H,F)]$ estimates a slower phase
than $F$. $\Rich[(M,H,F)]$ estimates a slower phase than both $F$ and
$\Rich[(L,M,H,F)]$, but the extrapolation might be more biased by the use
of three resolutions instead of four. The positive differences 
$\Delta\phi(F-\Rich[(M,H,F)])$ are slightly smaller than the positive 
$\Delta\phi(\Rich[(L,M,H,F)] - \Rich[(M,H,F)])$.
Although not shown in the figure, we verified that the Richardson
extrapolation that uses only two datasets is \emph{not} robust; the
result depends heavily on the choice of the pair of the datasets and
extrapolation results are incompatible with each other. We recommend
the use of at least three, and possibly four, datasets at
well-separated resolutions. 

These results indicate that the extrapolations are robust. We
choose $\Rich[(L,M,H,F)]$ as best data. The uncertainty of the best
data can be assigned either as
$\delta\phi=\Delta\phi(F-\Rich[(L,M,H,F)])$ or as 
$\delta\phi=\Delta\phi(\Rich[(L,M,H,F)]-\Rich[(M,H,F)])$. The
former choice gives the smallest error bars, and is incompatible
with the $\Rich[(M,H,F)]$. The latter choice instead assume
the  three-resolutions extrapolation $\Rich[(M,H,F)]$ is robust and
assigns a slightly more conservative error bar. 

The total error budget of SLy135135\_0060 is reported 
in Fig.~\ref{fig:bns_errbudget_sly}. The best approximation to 
the continuum data is given by $\Rich[(L,M,H,F)]$, with the uncertainty due to
truncation error calculated as
$\delta\phi_{(h)}=\Delta\phi(\Rich[(L,M,H,F)]-\Rich[(M,H,F)])$ as
discussed above.
The uncertainty $\delta\phi_{(r)}$ due to finite radius is instead
computed as described above using the polynomial extrapolation
with $K=2$ and last radius $r=1000$. The total error bar is computed as 
$\delta\phi=(\delta\phi_{(h)}^2 + \delta\phi_{(r)}^2)^{1/2}$ and shown
as a shaded area.

The use of the HO-Hyb-WENO scheme at these resolutions, combined with
the two extrapolation techniques described above, gives maximum phase
uncertainties $\delta\phi_{(h)}\lesssim0.05$ rad for $u\leq
u_\text{mrg}$.  The error budget is essentially flat at the level
$\delta\phi \sim \delta\phi_{(r)}\lesssim0.1$ rad, and dominated by 
finite-extraction errors (compare with
\cite{Bernuzzi:2011aq}).   

\begin{figure*}[t]
  \centering    
  \includegraphics[width=1\textwidth]{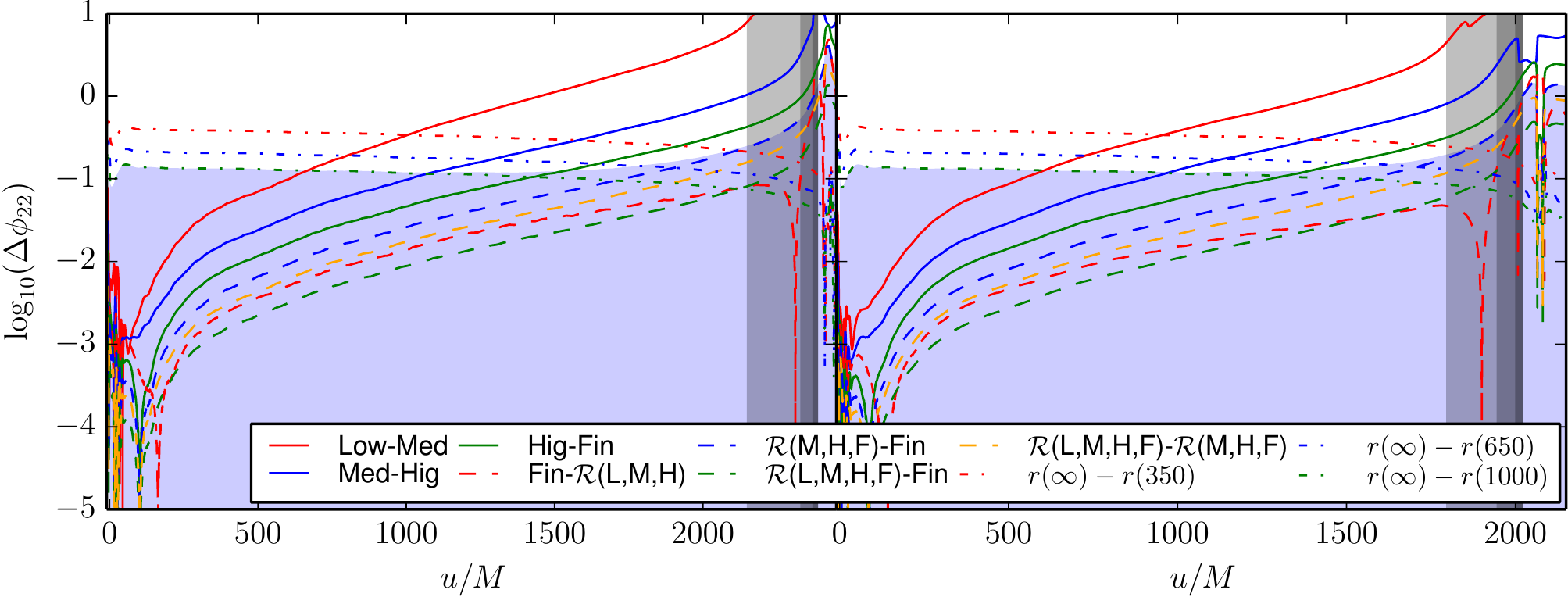}
  \caption{Phase error budget of SLy135135\_0038 (left) and 
    MS1b135135\_0038 (right) similar to Fig.~\ref{fig:bns_errbudget_sly}. Solid lines are
    phase differences between runs at different resolutions; dashed
    lines are differences with the Richardson extrapolated data;
    dashed dotted lines are differences between phase extracted at
    finite radii and the radius-extrapolated one. 
    The conservative error estimate described in the text is given by the blue shaded region.}
  \label{fig:bns_errbudget_10}
\end{figure*}

\subsection{SLy135135\_0038 and MS1b135135\_0038}

The error budgets for SLy135135\_0038 and MS1b135135\_0038 are
computed in the same way as above. The results are shown in
Fig.~\ref{fig:bns_errbudget_10}, and focus on the simulations with the
HO-Hyb-WENOZ scheme.
In these long runs, the finite-radius uncertainties are essentially
the same as those discussed for SLy135135\_0060. From
Fig.~\ref{fig:bns_errbudget_10} one observes that
$\delta\phi_{(r)}\lesssim0.1$ rad and slightly decreases towards
merger. The error bar is dominated by these finite-extraction errors
at early times, $u\lesssim1500~M$ for MS1b135135\_0038 and 
$u\lesssim1750~M$ for SLy135135\_0038. The main difference with
respect to 
 SLy135135\_0060 is related to the truncation error, and it is discussed below.

Considering the Richardson extrapolations, we find that the
extrapolations $\Rich[(L,M,H)]$ and $\Rich[(L,M,H,F)]$ are
effective and rather close to the 
data of grid $F$ (overestimating and underestimating the latter, respectively). 
Again, the positive difference $\Delta\phi = H - \Rich[(L,M,H)]$ is larger 
than the differences $\Delta\phi = \Rich[(L,M,H)] -F$ and
$\Delta\phi = F - \Rich[(L,M,H,F)]$, possibly
indicating a convergent behaviour of the extrapolation. However, the
difference $\Delta\phi = \Rich[(L,M,H)-F]$ shows a
zero crossing close to $u_\text{mrg}$ ($F$ data), likely due to the
very different merger time in the $L$ data with respect to other
datasets. The zero-crossing is not there in the difference $\Delta\phi
= F - \Rich[(L,M,H,F)]$, which is also the smallest difference
between grid $F$ and the various extrapolated data.
The positive difference $\Delta\phi=F-\Rich[(M,H,F)]$ is instead
larger than the differences with the extrapolations that include the
grid $L$. Because the $L$ data show overconvergence at late
time, a \textit{conservative} choice for our best data and uncertainty
is $\Rich[(M,H,F)]$ and $\delta_{(h)}\phi=F-\Rich[(M,H,F)]$. With
this choice we find that truncation errors dominate the error budget
at $u\gtrsim1500~M$, i.e. during the last $5.5$ GW cycles, for 
MS1b135135\_0038 and $u\gtrsim1750~M$, i.e. during the last $7$ GW cycles, for 
SLy135135\_0038. The maximum
error is reached at $u_\text{mrg}$ and $\delta\phi_{(h)}\sim1$~rad.
A more optimistic choice, though not fully reliable based on the convergence
result, is $\delta_{(h)}\phi=\Rich[(L,M,H,F)-\Rich[(M,H,F)]$
which leads to a maximum uncertainty of $\delta\phi_{(h)}\sim0.7-0.9$~rad.

Finally, we compare the new HO-Hyb-WENOZ extrapolated waveforms with our
previous simulations employing the LLF \cite{Bernuzzi:2014owa}. For
SLy135135\_0038 we find that the difference
$\Delta\phi=\phi({\rm LLF-WENOZ})-\phi({\rm HO-Hyb-WENOZ})\sim-1$ rad at
$u_\text{mrg}$, and is thus compatible with the error bar of the
HO-Hyb-WENOZ.
For MS1b135135\_0038 $\Delta\phi=\phi({\rm LLF-WENOZ})-\phi({\rm
  HO-Hyb-WENOZ})\sim+1.2$ rad at  
$u_\text{mrg}$, which is compatible with the error bar assigned to the
HO-Hyb-WENOZ and to the LLF data. In this case, the use of
HO-Hyb-WENOZ reduces by about a factor 3 the error estimated on the
phase.

\section{Summary}
\label{sec:conc}

We explored the use of finite-differencing high-order WENO schemes in
BNS inspiral-merger simulations, and compared those results with a
second-order (LLF) scheme employing the same reconstruction
methods. Simulations were performed at typical resolutions of the 3D
grid. Our findings are summarised in the following. 

(i) WENO methods are robust for these simulations; best results are
obtained with the HO-Hyb-WENOZ scheme which significantly improves
over the LLF, see e.g. Fig.~\ref{fig:bns_phiconv_sly}.
The HO-Hyb-WENOZ scheme allows us to consistently measure
self-convergence in all the different datasets (triplets); and, hence,
to robustly build an error budget by extrapolating consistently the
finite-resolution datasets;

(ii) At the considered resolutions, the high-order convergence rate
cannot be obtained due to nonoptimal values of the WENO weights. We
experimentally observe that truncation errors scale at second-order
rate for both single star and binary spacetimes. We do not align
waveforms in time and/or phase in these analysis.
In three orbits simulations (SLy135135\_0060), convergence is observed
by resolving the 
NSs with $\sim64^3$ grid points.
In ten orbits simulations, convergence is observed by resolving the
NSs with at least $\sim96^3$ grid points;

(iii) The HO-Hyb-WENOZ scheme combined with the Richardson
extrapolation allows one to compute a robust error budget.   
Notably, our error budget procedure does not involve waveform alignment.
The error budget for the three orbits runs (SLy135135\_0060) is dominated by
finite-extraction uncertainties, see Fig.~\ref{fig:bns_errbudget_sly}. The overall
phase uncertainty accumulated to merger ($u=u_\text{mrg}$) is below $0.1$ rad.
The error budget of $\sim$ ten orbits runs is dominated by truncation
error during the last 3-4 orbits; the phase uncertainty grows from
$0.1$ to a maximum of $\sim1$ rad at merger, see Fig.~\ref{fig:bns_errbudget_10};

(iv) Comparing our results with
\cite{Radice:2013hxh,Radice:2015nva,Radice:2016gym}, we find that,
overall, our conclusions are in line with previous
work. Common features are the following: the robustness of the high-order finite
differecing scheme, the fact that the formal high-order accuracy is
not achieved, and the magnitude of the phase uncertainties. the main differences
are the performances of the MP5 reconstruction and the observed
convergence order ($\sim3$ instead of $2$). Both differences might be 
due to the use of different basic flux formula in the HO algorithm
(the Roe solver instead of the LLF) and/or other implementation
details, notably the use of positivity preserving limiters and the
atmosphere treatment \cite{Radice:2013xpa}. 
Future work on the BAM code could be devoted to design improved and
specific WENO weights for the problem.

We conclude that our high-order WENO implementation can be efficiently
used for high-quality waveform production, and in future large-scale
investigations of the binary parameter space. 
The computational cost for a $\sim10$ orbit simulation composed of
four resolutions is about $\sim 650$k core hours. 
We expect that the maximum phase uncertainty can be further reduced below
  $\delta\phi\lesssim0.5$ rad by simulating an additional resolution at 
  $\sim192^3$ grid points. Using these resolutions, the BAM's parallel
efficiency is $\sim80\%$ on 1024 processing units (strong scaling
tests), with a simulation speed of $\sim15M_\odot$/hour.
The computational cost for these five-resolutions is estimated as $\sim
1$M core hours. Several BNS configurations could be simulated using a
large-scale HPC allocation.

\begin{acknowledgments}
  It is a pleasure to thank Bernd Br\"ugmann and members of the Jena
  group, including David Hilditch and Marcus Thierfelder, for many 
  discussions and help with the BAM code. 
  This work started at Jena and was supported in part by DFG grant
  SFB/Transregio~7 ``Gravitational Wave Astronomy''. 
  We also thank Andrea Mignone for clarifications, and for sharing some of
  the routines of the PLUTO code; David Radice for interactions and
  comments on the manuscript; and Maximiliano Ujevic for computing the
  initial data employed in this article.  
  Computations where performed on LRZ (Munich), Juropa (J\"ulich),
  Stampede (Texas, XSEDE allocation), and the Jena group local cluster. 
\end{acknowledgments}

\appendix

\begin{figure*}[t]
  \centering
  \includegraphics[width=0.49\textwidth]{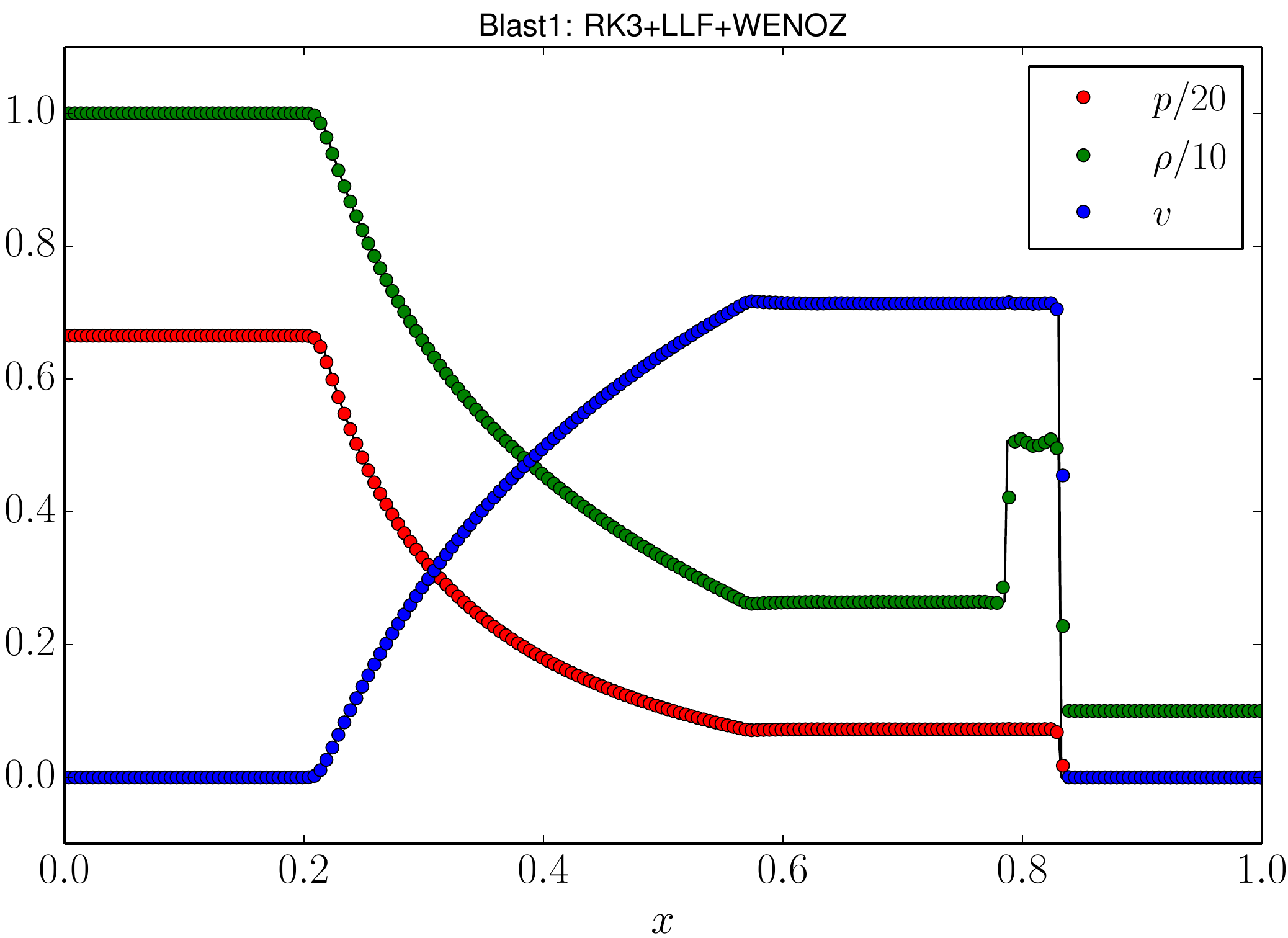}
  \includegraphics[width=0.49\textwidth]{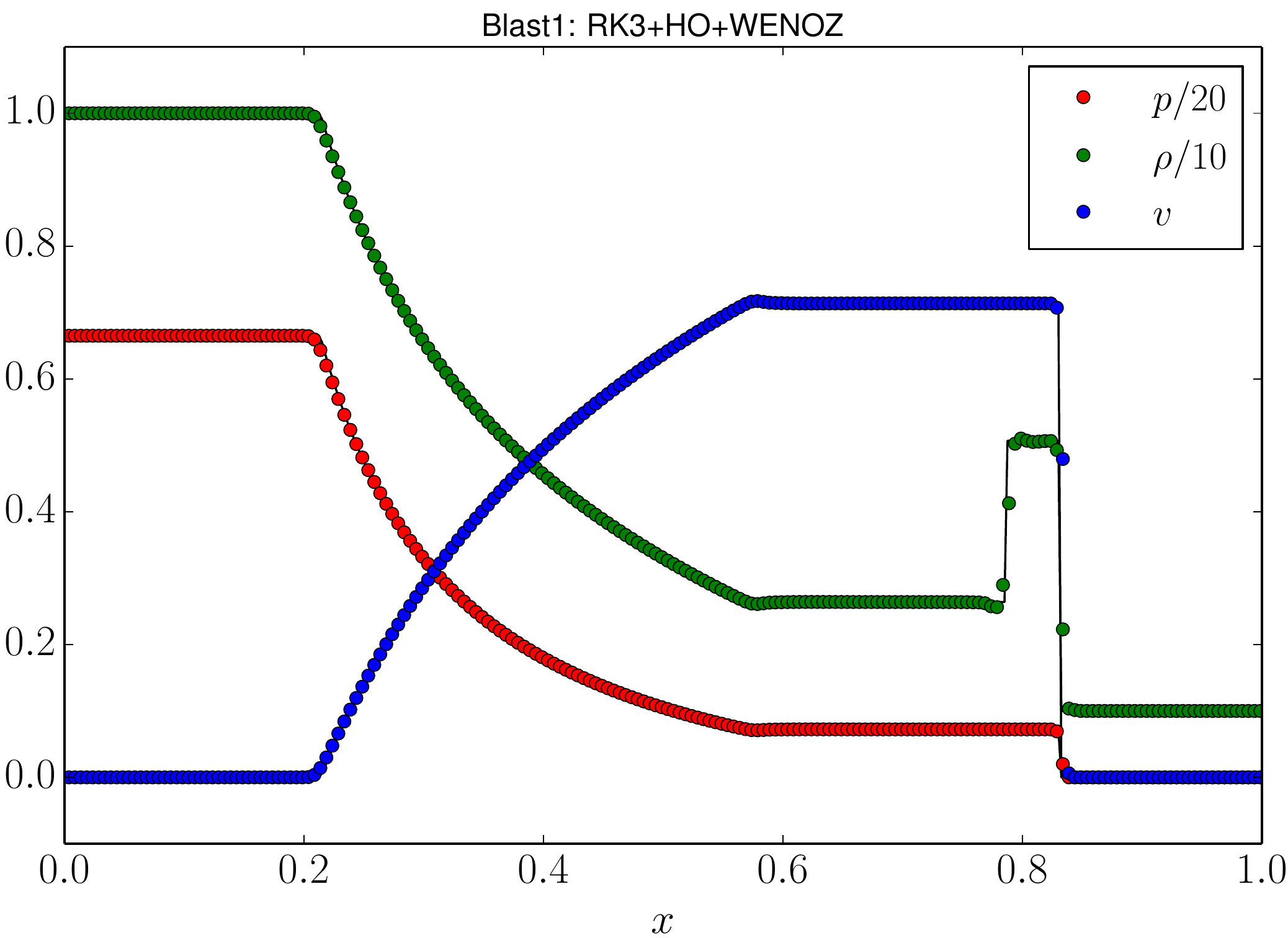}
  \includegraphics[width=0.49\textwidth]{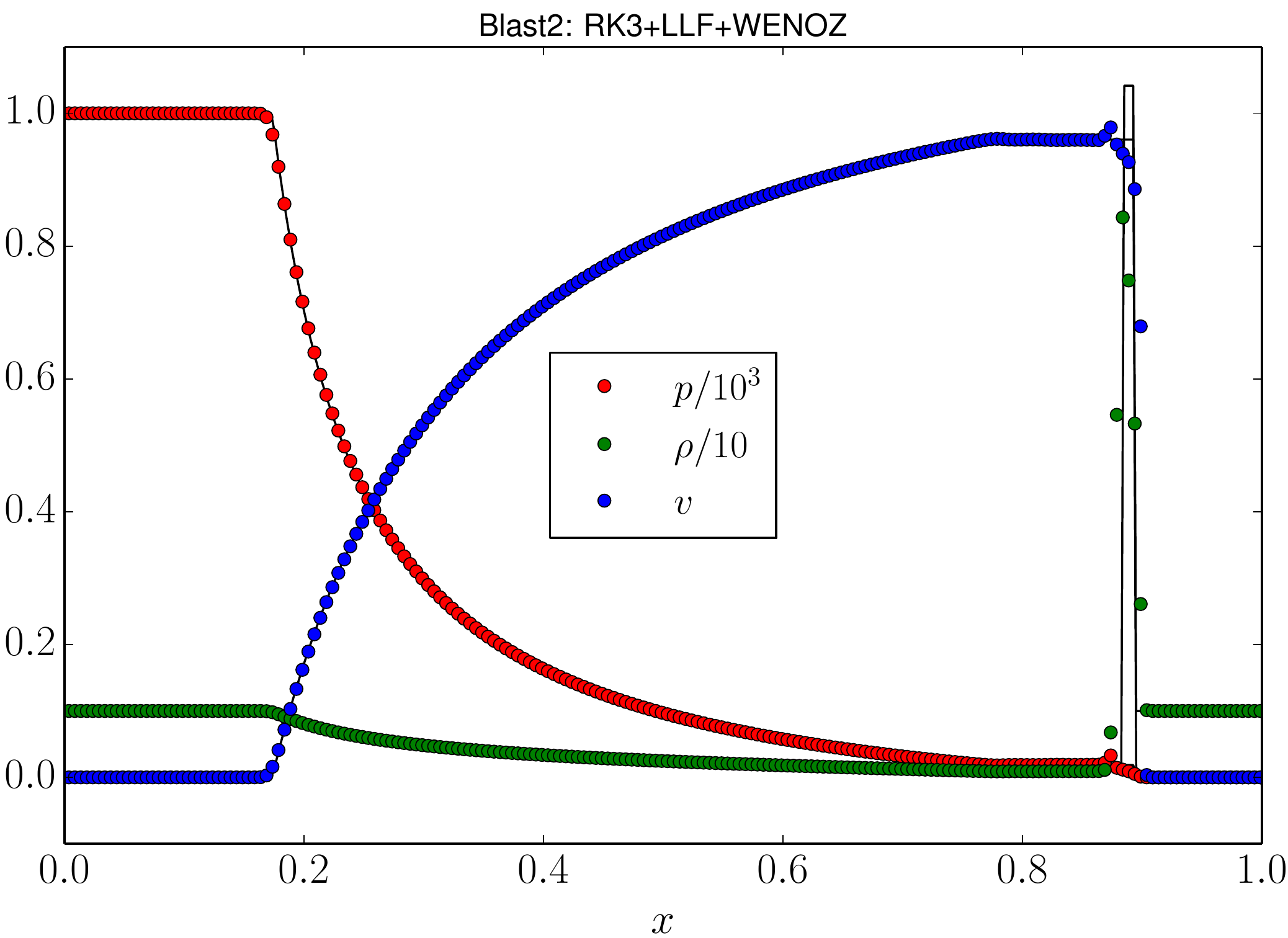}
  \includegraphics[width=0.49\textwidth]{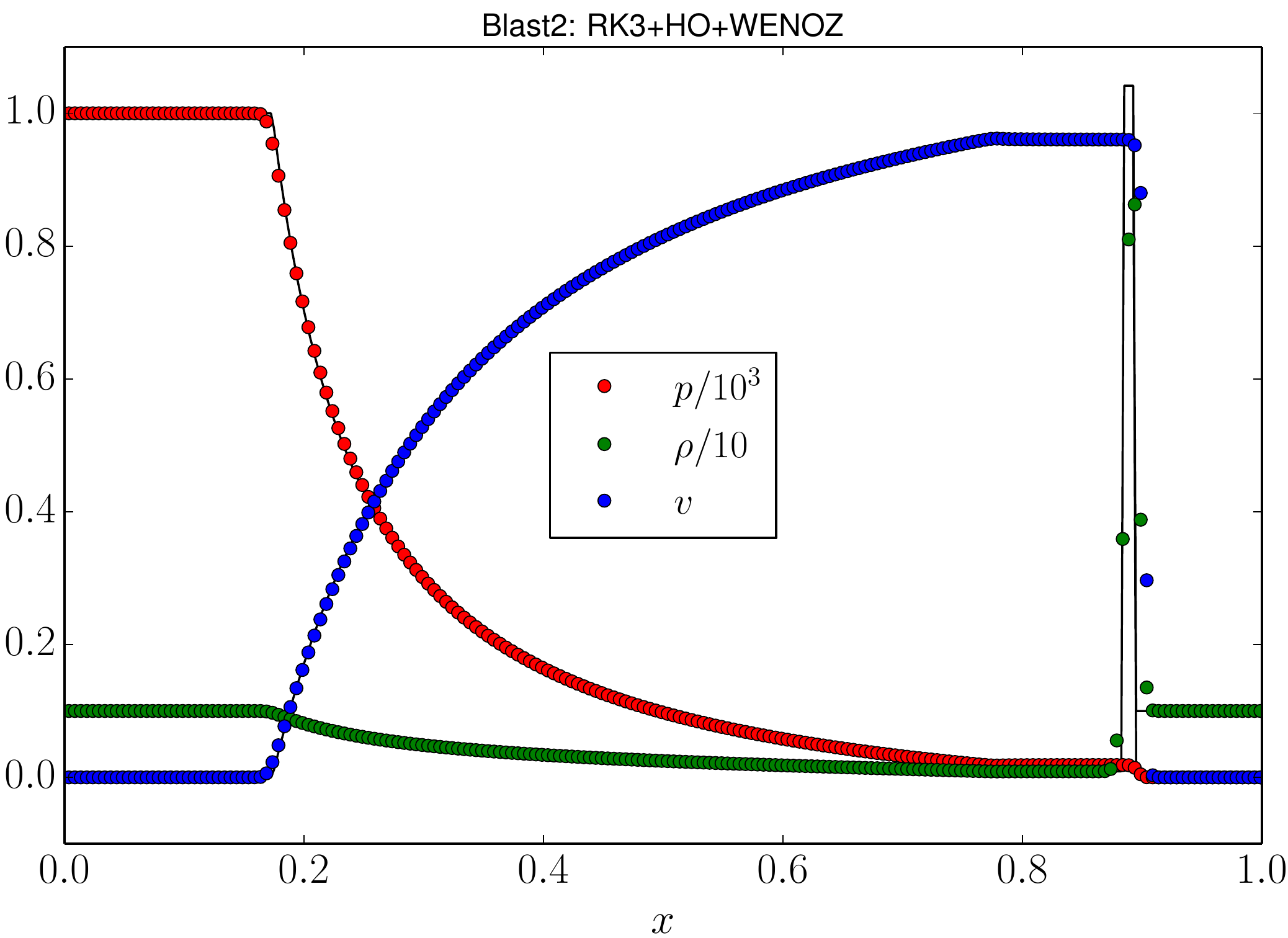} 
    \caption{Relativistic blast waves solutions at $t=0.4$ with
      $n=400$ points. Solid lines are the exact solutions.} 
    \label{fig:blast} 
\end{figure*}

\begin{figure}[t]
  \centering
\includegraphics[width=0.49\textwidth]{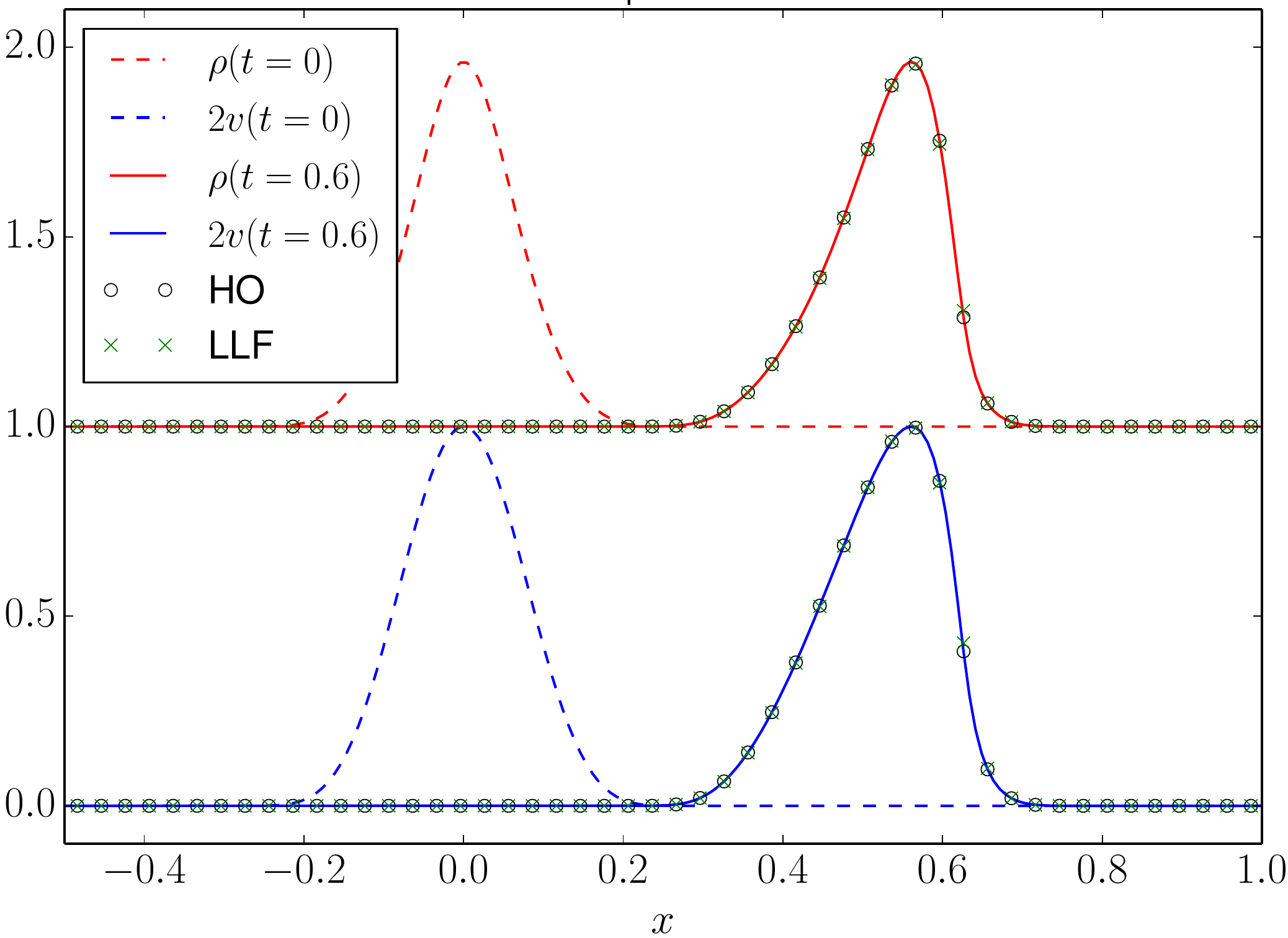}
\includegraphics[width=0.49\textwidth]{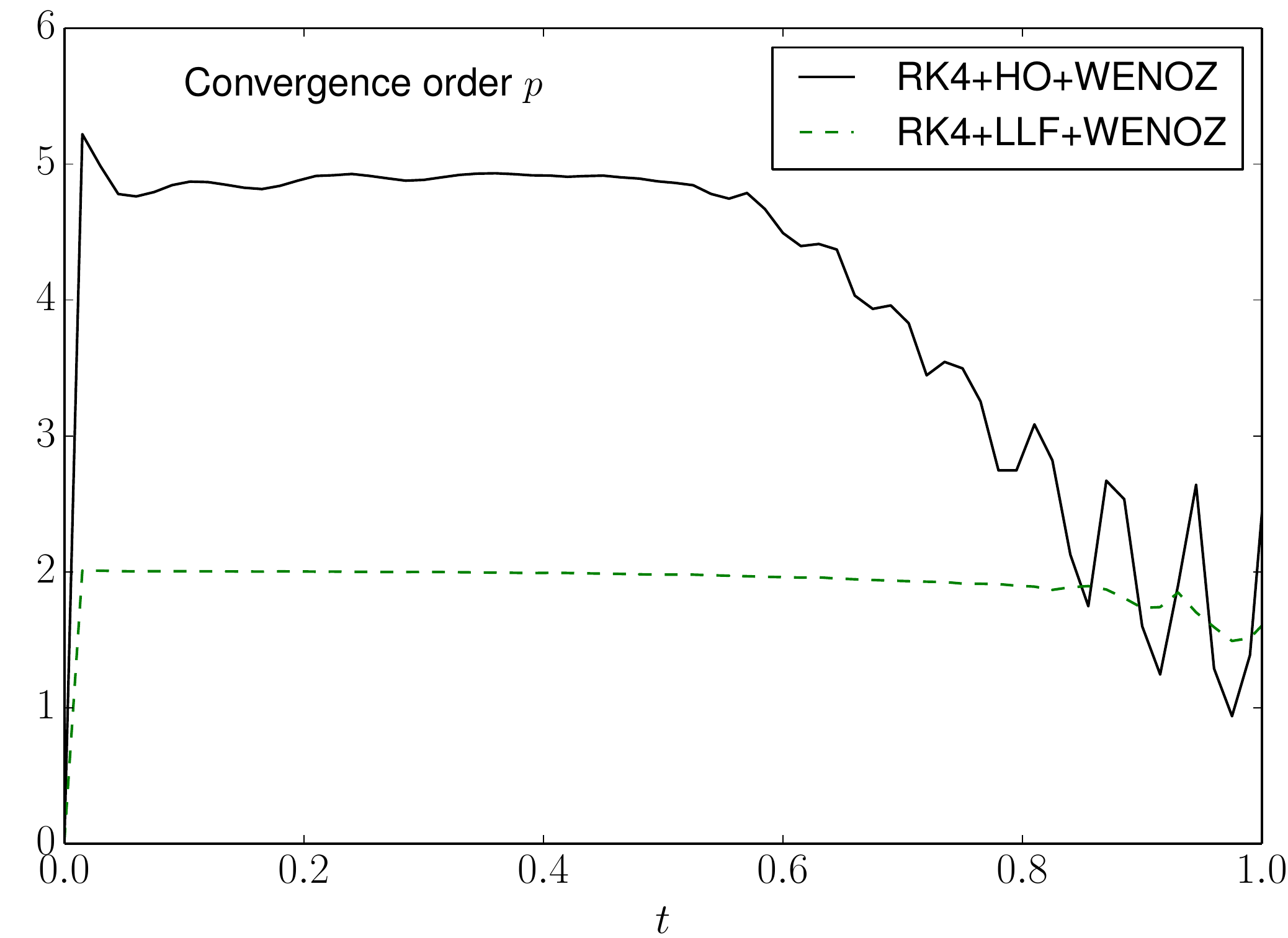}
    \caption{Relativistic simple wave. Top: solution with $n=400$ points 
     (shown $200$) and CFL factor $0.125$. Bottom: evolution of the
      experimental self-convergence factor as computed from numerical
      data at resolutions $(400,800,1600)$ points.} 
    \label{fig:simplewave} 
\end{figure}

\section{Flat Spacetime 1D Tests}
\label{app:flat_test}

\begin{table}[t]
  \centering    
  \caption{ \label{tab:simplewave_conv} Convergence results for the 1D
    simple wave test at $t=0.6$. The convergence rate $p$ is
  calculated dividing successive $L_1$ norm values.}
  \begin{tabular}{ccccccc}        
    \hline
    \hline
    Scheme & $n$ & $L_1$ & $p$ & $L_2$ & $p$\\
    \hline
    LLF-WENOZ & 200 & 2.271e-03 & - & 6.431e-03& -\\
    & 400 & 5.974e-04& 1.927 &1.810e-03 &1.829\\
    & 800 & 1.521e-04 &1.974 &4.657e-04 &1.959\\
    & 1600 & 3.828e-05 &1.990 &1.174e-04 &1.988\\
    & 3200 & 3.209e-06 &1.997 &1.698e-05 &1.996\\
    \hline
    HO-WENOZ & 200 & 5.988e-04& - &1.790e-03 & -\\
    &400 & 4.059e-05 &3.883 &1.512e-04 &3.566\\
    &800 & 1.677e-06 &4.597 &7.470e-06 &4.339\\
    &1600 & 5.623e-08 &4.898 &2.567e-07 &4.863\\
    &3200 & 6.089e-10 &4.947 &4.621e-09 &5.002\\
    \hline
    LLF-MP5 & 200   & 7.579e-04 & - & 3.713e-03& -\\
            & 400   & 1.991e-04 & 1.928 &1.043e-03&1.832\\
            & 800   & 5.081e-05 & 1.970 &2.685e-04&1.957\\
            & 1600  & 1.281e-05 & 1.988 &6.772e-05&1.987\\
            & 3200  & 3.209e-06 & 1.997 &1.698e-05&1.996\\
    \hline
    HO-MP5   & 200    & 2.774e-04 & -     & 1.466e-03 & -\\
             & 400    & 1.365e-05 & 4.346 & 9.129e-05 & 4.005 \\
             & 800    & 5.455e-07 & 4.645 & 4.334e-06 & 4.396 \\
             & 1600   & 1.797e-08 & 4.924  &1.487e-07 & 4.866 \\
             & 3200   & 5.555e-10 & 5.016 & 4.576e-09 & 5.022\\
    \hline 
    \hline
 \end{tabular}
\end{table}

The implementation of the HO schemes in BAM has been tested against standard
special relativistic hydrodynamics benchmarks. In particular, we
present here the two 1D relativistic blast wave tests described
in~\cite{Marti:1999wi} and a 1D evolution of isentropic smooth
waves~\cite{Liang:1977ApJ...211..361L,Anile:1990rfmf.book.....A}
(see also~\cite{Colella2006347,Zhang:2005qy,Radice:2012cu,Bugner:2015gqa} for similar
tests). 1D benchmarks are important because exact solutions
are available in these cases and, as already mentioned, the multi-D scheme
is obtained by successive application of 1D procedures. 
In practice, only new 1D routines has been implemented for the HO 
algorithm. 

The blast wave tests assess that the numerical scheme is
nonoscillatory, handles solutions with shocks properly, and captures 
correctly all the elementary waves of the evolution of the Riemann
problems. The evolution of simple waves is instead used to prove the
scheme achieves high-order convergence for smooth flow solutions. 
For comparison, all the tests were also run with the second-order LLF
scheme based on primitive reconstruction employing the same
WENOZ or MP5 reconstruction as in the HO case. 
Let us stress that in the HO and LLF algorithms the same
reconstruction is applied to different 
quantities: either projected fluxes or primitives variables.

Initial data for the relativistic blast waves are set exactly as
described in~\cite{Marti:1999wi} (Problem 1 and 2), and a
$\Gamma=5/3$ EOS is employed. The numerical solutions at $t=0.4$ are
shown in Fig.~\ref{fig:blast}. They are
computed with $400$ grid points in the domain $[0,1]$ (resolution $h=0.0025$).
The RK3 time integrator and CFL 
factor $0.25$ are used. The exact solution is also plotted as solid lines. As
evident from the figures, the simulations 
reproduce all the features of the exact solution, and no relevant
differences are found between the HO and the LLF scheme. (Note, however,
a small difference in the resolution of the shock wave.)

Relativistic simple waves are nonlinear elementary waves analogues of
planar acoustic waves~\cite{Liang:1977ApJ...211..361L,Anile:1990rfmf.book.....A}. The exact solutions 
can be found implicitly by the method of characteristics. During
the evolution, shock formation is expected in a finite time due to
nonlinearities. Right-propagating simple wave 
initial data are set up by (i)~choosing a reference state ($\rho=1$,
$v=0$), (ii)~prescribing a velocity perturbation, and (iii) computing the
sound speed according to the value of the Riemann invariant
(Eq.~(II.15) of~\cite{Liang:1977ApJ...211..361L}). The other quantities follow from
the EOS. The velocity profile is here set as, 
\be
\label{eq:simplewave_v}
v  = a \ \Theta\left(|x|-X\right) \ \sin^6\left( \frac{\pi}{2}
\left(\frac{x}{X} - 1\right) \right) \ , 
\ee
where $\Theta(x)$ is the Heaviside function, $a=0.5$ and
$X=0.3$. Assuming a polytropic EOS with $\Gamma=5/3$ and $K=100$, the
value of the sound speed in the reference state is
$c_{s,0}\simeq0.815$; the velocity and density initial profiles are
shown in Fig.~\ref{fig:simplewave} (top).
Note that the profile is smooth enough to guarantee that the WENO
weights are close to optimal.
Numerical solutions are computed on the domain
$[-1.5,1.5]$, with the RK4 integrator and a CFL factor of $0.125$.
During the evolution the initial profiles  progressively steepen and
finally a shock is formed~\footnote{%
  Shock formation is estimated using Eq.~(III.2)
  of~\cite{Liang:1977ApJ...211..361L} with the velocity profile
  specified by~\eqref{eq:simplewave_v}.}
at around $t\simeq0.63$. Figure~\ref{fig:simplewave}  
(top) shows the numerical solutions at $t=0.9$ for a resolution of
$400$ points ($h=0.0075$). The exact solution
is also plotted as solid lines. As evident, the simulations reproduce the
correct physics. The formal convergence order of the numerical scheme
can be checked at early times (before the shock forms).
The bottom panel of Fig.~\ref{fig:simplewave} shows the self-convergence
factor of a three-levels convergence test.
In Tab.~\ref{tab:simplewave_conv} we report the simulation errors as
function in the $L_1$ and $L_2$ norms, at different resolutions and
$t=0.6$. Both schemes converge to the exact solution at the expected
rate for sufficiently high resolutions.

\section{Extrapolation in resolution alternative to the Richardson method}
\label{app:extr2}

Ref.~\cite{Hotokezaka:2015xka} proposes a method for the resolution
extrapolation that is alternative to the Richardson.
In this appendix we apply that method to our 
SLy135135\_006 data obtained with HOHyb-WENOZ scheme, and compare the
results with those in Sec.~\ref{sec:err}.  

Low resolution simulations are more affected by numerical dissipation than 
high resolution simulations. This results in earlier merger times
for decreasing  
resolution. As a consequence, Ref.~\cite{Hotokezaka:2015xka} argues
for the
necessity to compensate for this effect by rescaling waveforms in
time, $u \rightarrow \eta u$, and phase, $\phi \rightarrow \eta \phi$.
The parameter $\eta$ is determined by minimizing 
\begin{align}
I_{2,1} = \min_{\eta',\Phi} \int_{t_i}^{t_f} \text{d}u [ A_{2,22}
  (\eta' u) e^{i ( \eta' \phi_2(\eta' u) + \Phi)}  \nonumber \\  
  - A_{1,22}(u) e^{i \phi_1(\eta)}]^2,
\end{align}
where $A_{1,22},\phi_{{1,22}}$ refer to the amplitude
 and phase of the best resolved run and
 $A_{2,22},\phi_{{2,22}}$ to the other resolutions.  
We obtain the following values
$\eta_{\rm L,F} = 0.98651$,
$\eta_{\rm M,F} = 0.99584$,
$\eta_{\rm H,F} = 0.99867$.
As in~\cite{Hotokezaka:2015xka} we estimate the convergence order $p$ from a 
three-level self-convergence test, for $\eta_{\rm M,F}, \eta_{\rm H,F}$ we  
obtain $p=1.96$ and for $\eta_{\rm L,F}, \eta_{\rm M,F}$ $p=2.34$. 
This allows us to scale the H-resolution data 
with a scaling parameter of $\eta_{\rm F,\infty} = 0.99758$. 

\begin{figure}[t]
  \centering
    \includegraphics[width=0.49\textwidth]{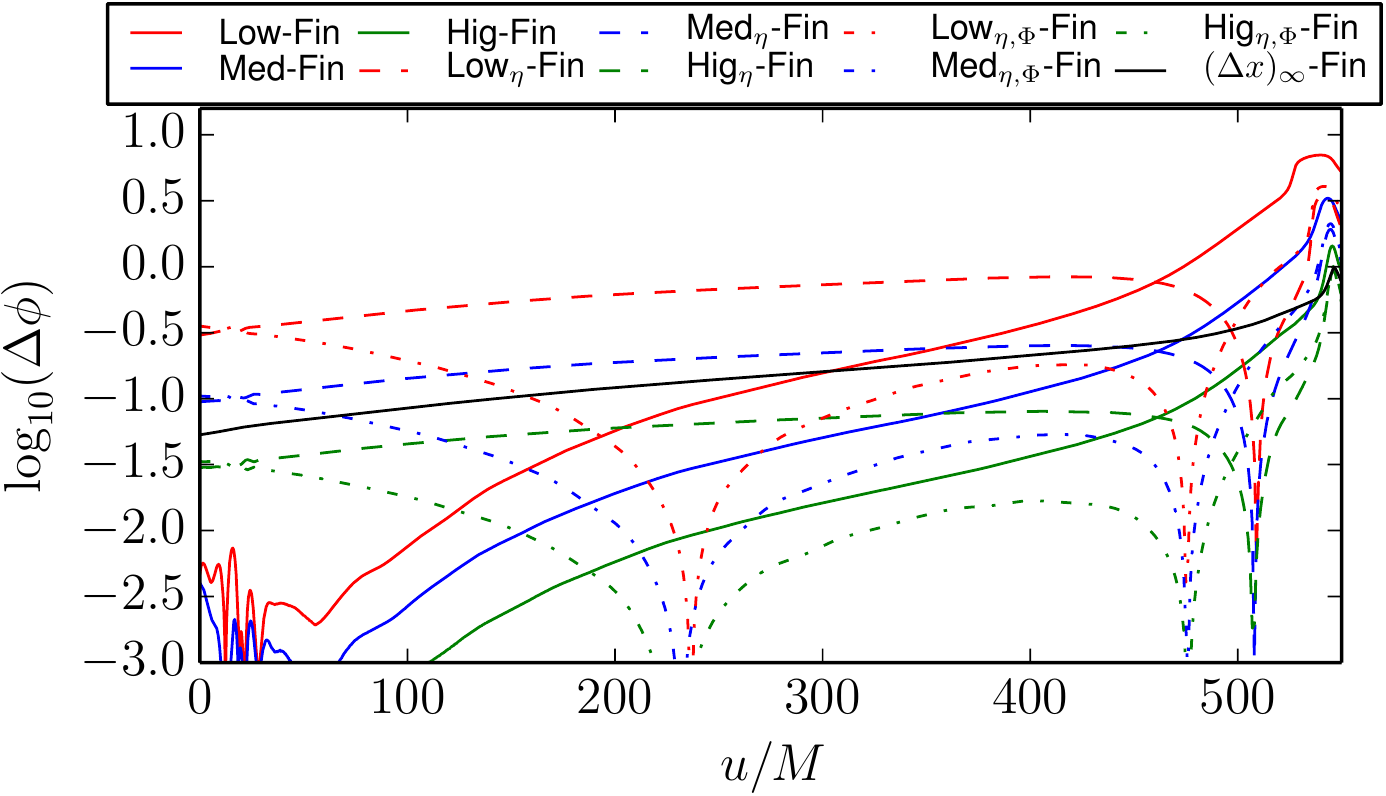}
    \caption{Phase differences for Sly135135\_006. Solid lines refer to the phase differences 
    of the original data (L,M,H) to the highest resolved simulation (F). 
    Dashed lines correspond to the phase difference between $\eta$ rescaled data 
    and the original F-simulation. Dash-dotted lines are $\eta,\Phi$-rescaled data compared 
    to H-resolution. The solid black line marks the difference of the extrapolated data to 
    infinite resolution according to the text and F-resolution.}
    \label{fig:phase_shibata} 
\end{figure}

In Fig.~\ref{fig:phase_shibata} we compute the phase difference between 
the F-resolution and the (L,M,H)-simulations without shift, with 
$\eta$-shift, and with $\eta,\phi$-shift. We also include the phase
difference between F-data and the waveform obtained with the
extrapolation described above. 
We observe three main effects of the rescaling: 
(i) the phase difference during the beginning of the simulation increases;
(ii) around $u_\text{mrg}$ the phase differences are smaller than for the nonshifted setup;
(iii) we see zero crossing caused by the rescaling procedure.

Following~\cite{Hotokezaka:2015xka} the uncertainty is 
estimated by the relative uncertainty of the merger time 
multiplied by total phase of the simulation, i.e. $\Delta \phi = \phi(u_{\rm mrg}) \; \Delta u_{\rm mrg}/u_{\rm mrg}$. 
For SLy135135\_006, we have $\phi(u_{\rm mrg})\approx 40.8$~rad,
$u_{\rm mrg}\approx504M$. 
The difference $\Delta u$ is estimated either
from the difference between the F-resolution and 
the extrapolated data as $\Delta u^{\rm mrg} = 0.9M$, or based on the 
uncertainty of $p$ as extracted from the two different triplets to
$\Delta u^{\rm mrg}= 0.3M$, cf.~Ref.~\cite{Hotokezaka:2015xka}. 
Consequently the total uncertainty lies between $0.07$ and 
$0.025$~rad for two choices respectively. 
The optimistic uncertainty is a factor of $\sim 2$ smaller than the
error assigned from the Richardson extrapolation. 
Because of the zero crossings, and the fact that the error budget 
strongly depends on the convergence order estimated solely from
the merger time (at which the uncertainties are maximal), we prefer to
use the simpler method outlined in Sec.~\ref{sec:err}.

%

\end{document}